\documentclass{pasa}%

\title[{\sc Profiler} -- Galaxy Light Profile Decomposition]{{\sc Profiler} -- A Fast and Versatile New 
Program for Decomposing Galaxy Light Profiles}
\author[B. C. Ciambur]{Bogdan C. Ciambur$^1$\thanks{E-mail: bciambur@swin.edu.au}\\
\affil{$^1$Centre for Astrophysics and Supercomputing, Swinburne University of Technology, Hawthorn, VIC 3122, Australia}}
\jid{PASA}

\usepackage{aas_macros}
\usepackage{wrapfig}
\usepackage{lscape}
\usepackage{rotating}
\usepackage{epstopdf}
\usepackage{mathrsfs,amssymb,amstext}
\usepackage{amsmath}
\usepackage{amsfonts}
\usepackage{mathrsfs}
\usepackage{euscript}
\usepackage{hyperref} 
\usepackage{color}
\usepackage{float}
\usepackage{natbib}
\hypersetup{colorlinks,citecolor=blue,linkcolor=blue,urlcolor=blue}
\newcommand{\prof}{{\sc Profiler}}

\usepackage{rotating}

\begin{document}%
\begin{abstract}

I introduce \prof, a new, user-friendly program written in {\sc Python} and designed to analyse the radial surface brightness profiles of galaxies. 
With an intuitive graphical user interface, \prof~can accurately model a wide range of galaxies and galaxy components, such as elliptical galaxies, the bulges of spiral and lenticular galaxies, nuclear sources, discs, bars, rings, spiral arms, etc., with a variety of parametric functions routinely employed in the field (S\'ersic, core-S\'ersic, exponential, Gaussian, Moffat and Ferrers). In addition to these, \prof~can employ the broken exponential model (relevant for disc truncations or antitruncations) and two special cases of the edge-on disc model: namely along the major axis (in the disc plane) and along the minor axis (perpendicular to the disc plane). \prof~is optimised to work with galaxy light profiles obtained from isophotal measurements which capture radial gradients in the ellipticity, position angle and Fourier harmonic profiles of the isophotes, and are thus often better at capturing the total light than two-dimensional image-fitting programs. Additionally, the one-dimensional approach is generally less computationally expensive and more stable. In \prof, the convolution of either circular or elliptical models with the point spread function is performed in two-dimensions, and offers a choice between Gaussian, Moffat or a user-provided data vector (a table of intensity values as a function of radius) for the point spread function. I demonstrate \prof's features and operation by decomposing three case-study galaxies: the cored elliptical galaxy NGC~3348,  the nucleated dwarf Seyfert I galaxy Pox~52, and NGC~2549, a structurally complex, double-barred galaxy which additionally displays a Type II truncated disc viewed edge-on.

\prof~is freely available at \url{https://github.com/BogdanCiambur/PROFILER}.
\end{abstract}
\maketitle%
\section{INTRODUCTION}
\label{sec:intro}

Galaxies are complex structures assembled through a variety of physical processes which act at different stages of their life (e.g., gas accretion; star formation; disc formation, growth and buckling; bar formation and buckling etc.; as well as mergers and interactions with neighbouring galaxies). The result is a rich variety of galactic components in the observed galaxy population. Classifying galaxies based on these structures, in the optical and/or near-infrared bands has been and still is now common practice (e.g., \citealt{Jeans1919}, \citealt{Hubble1926}, \citealt{deVaucouleurs1959}, \citealt{Sandage1975}, \citealt{deVaucouleurs+1991}, \citealt{Abraham+2003}, \citealt{Buta+2015}, etc.).

A quantitative structural classification requires a reliable method to separate out each structural component from the others that make up the galaxy. Moreover, individually analysing each constituent probes the specific physical or dynamical processes associated with it and thus provides insight into galaxy evolution. The common practice is to model one of the fundamental diagnostics of a galaxy's structure, namely its radial light (or surface brightness) profile (SBP), by decomposing it into a sum of analytical functions, with each function representing a single component (e.g., \citealt{Prieto+2001}, \citealt{Balcells+2003}, \citealt{Blanton+2003}, \citealt{Naab+2006}, \citealt{Graham&Worley2008}, etc.; the reader will find an insightful and comprehensive review of the long history of modelling galaxy light profiles in \citealt{Graham2013}).

Galaxy SBPs are commonly extracted from images by fitting quasi-elliptical isophotes as a function of increasing distance (semi-major axis) from the photometric centre of the galaxy. In such schemes, the isophotes are free to change their axis ratio (ellipticity), position angle (PA) and shape (quantified through Fourier harmonics) with radius, which ensures that the models capture the galaxy light very well. A popular tool for this is the IRAF task {\sc Ellipse} (\citealt{Jedrzejewski1987}), which works well for galaxies whose isophotes display low-level deviations from pure ellipses (e.g., elliptical galaxies or disc galaxies viewed relatively face-on). For more complex isophotal structures however, (e.g., edge-on disc galaxies, X/peanut-shaped bulges, bars and barlenses) {\sc Ellipse} has been shown to fail and the newer IRAF task {\sc Isofit}\footnote{\url{https://github.com/BogdanCiambur/ISOFIT}} (\citealt{Ciambur2015}) is more appropriate.

A somewhat different approach to performing galaxy decomposition is to directly fit the galaxy's (projected) light distribution in 2D (i.e., the galaxy image). Recent years have seen the advent and development of a number of programs dedicated to this purpose, notably {\sc GIM2D} (\citealt{Simard+2002}), {\sc Budda} (\citealt{deSouza+2004}),  {\sc Galfit} (\citealt{Peng+2010}) and {\sc Imfit} (\citealt{Erwin2015}).

In support for the 2D method, \cite{Erwin2015} has invoked several drawbacks of 1D profile modelling, namely that it is unclear which azimuthal direction to model (major axis, minor axis or other), that most of the data from the image is discarded, and that non-axisymmetric components (such as bars) can be misinterpreted as axisymmetric components, and their properties cannot be extracted from a 1D light profile. While these issues certainly apply when one extracts the SBP by taking a 1D cut from a galaxy image, all of these issues are resolved if the SBP is obtained from an isophotal analysis. In particular, fitting isophotes makes use of the entire image (so no data are discarded) and apart from the SBP itself, this process additionally provides information about the isophotes' ellipticities, position angles (PAs), and deviations from ellipticity (in the form of Fourier modes). All of this information is sufficient to completely reconstruct the galaxy image for even highly complex and non-axisymmetric isophote shapes (see \citealt{Ciambur2015} and Section \ref{sec:ex} of this paper). Having these extra isophote parameters allows one to obtain the SBP along any azimuthal direction, and identify and quantitatively study non-axisymmetric components such as bars or even peanut/X--shaped bulges (\citealt{Ciambur+2016}). It is therefore recommended to always use isophote tables rather than image cuts in 1D decompositions.

Overall, both 1D and 2D decomposition techniques present benefits as well as disadvantages. The 2D image-modelling technique has the advantage that every pixel (except those deliberately masked out due to contaminating sources) in the image contributes directly to the fitting process, whereas in an isophotal (1D) analysis, pixels contribute in an azimuthal-average sense. Multicomponent systems with different photometric centres can also pose a problem for 1D SBPs, which assume a single centre for all components at $R=0$\footnote{This applies also to ring components, which have their brightest point at $R > 0$ along the 1D profile. This radial parameter represents the radius of the ring, while its centre is still assumed to be at $R = 0$.}, but can however be easily modelled in 2D. On the other hand, 2D codes suffer from the fact that each component has a single, fixed value for the ellipticity, PA, and Fourier moments (such as boxyness or discyness, and also higher orders), which can in some cases limit the method considerably. Triaxial ellipsoids viewed in projection can have radial gradients in their ellipticities and PAs (\citealt{Binney1978}, \citealt{Mihalas&Binney1981}), an effect captured in a 1D isophotal analysis (where both quantities can change with radius) but not in a 2D decomposition\footnote{Note, however, that the 2D code {\sc Imfit} can generate 2D images from line-of-sight integration of 3D luminosity density.}. 

There are notable examples in the literature where the 1D method has been preferred over the 2D technique. One such case is the decomposition of the {\sc Atlas$^{\rm 3D}$} (\citealt{Cappelari+2011}) sample{\footnote{\url{http://www-astro.physics.ox.ac.uk/atlas3d/}} of early-type galaxies, in \cite{Krajnovic+2013}. I point the reader to Sec. 2 and Appendix A of their paper, where they discuss both methods and test the performance of their preferred 1D method against a 2D analysis (with {\sc Galfit}).  Another illuminating example is in \cite{Savorgnan&Graham2016}. They performed both 1D (with private code) and 2D (with {\sc Imfit}) decompositions of 72 galaxies, out of which 41 did not converge or did not give meaningful solutions in 2D, whereas only 9 could not be modelled in 1D. Sec. 4.1 in their paper also provides an insightful and practical comparison between 1D and 2D galaxy modelling techniques. 

The past few decades have seen a flurry of 2D image-fitting codes, whereas publicly available tools that focus on 1D decompositions are scarce. In this paper I present \prof, a freely-available code written in {\sc Python} and designed to provide a fast, flexible, user-friendly and accurate platform for performing structural decompositions of galaxy surface brightness profiles. 

The remainder of the paper is structured as follows. In Section \ref{sec:input} I describe the input data and information required by \prof~prior to the decomposition process. Section \ref{sec:model} is a concise review of typical galaxy components and the analytical functions employed to model them. Section \ref{sec:fitting} then details the fitting process, and Section \ref{sec:ex} provides three example applications, each illustrating different features of \prof: modelling a core-S\'ersic galaxy, the user-provided PSF vector feature and modelling a structurally complex edge-on galaxy with a spheroid, two nested bars and a truncated disc. Finally, I summarise and conclude with Section \ref{sec:cons}.

\section{THE INPUT DATA}\label{sec:input}

With a view to streamline the decomposition process, \prof~has a built-in Graphical User Interface (GUI) coded in the standard {\sc Python} package {\sc TkInter}. This ensures that the decomposition process is entirely interactive, with all settings, options and input information readily changeable through buttons, text-box and ckeck-box widgets in the main GUI. Thus the need to perpetually change a separate configuration file each time one wishes to modify settings is eliminated, and the user can employ the vizualisation tools (which will be discussed in Section \ref{sec:ex}) and the GUI to make any required tweaks, until the solution is reached. Figure \ref{fig:profiler} presents the GUI, with most widgets active for illustration purposes (i.e., text-boxes contain default numbers or strings, and two components -- a S\'ersic and an exponential -- have been activated by pressing the corresponding buttons). Note that the user must specify all this galaxy-specific information on a case-by-case basis, as is detailed below.

\subsection{The Surface Brightness Profile}

\prof~was designed to work with isophote tables generated by either {\sc Ellipse} or {\sc Isofit} as, apart from the galaxy light profile itself, the two programs provide useful ancillary information, such as the isophotes' ellipticities, position angles, $B_4$ and $B_6$ harmonic amplitudes etc. The user can input data from both the above sources or, additionally, they can provide a simple table consisting of two columns, namely radius $R$ and intensity $I(R)$. 

Instrument-specific details are additionally required, in particular the CCD angular size of a pixel, in arcsec, and the zero-point magnitude. The isophote intensity $I$ is then converted into surface brightness $\mu$ (in magnitudes arcsec$^{-2}$) through:

\begin{equation}
\mu(R) = m_0 - 2.5{\rm log}_{10}\left[\frac{I(R)}{ps^2}\right],
\label{equ:sb}
\end{equation}

where $m_0$ is the zero-point magnitude and $ps$ is the pixel angular size. In Equation \ref{equ:sb}, $R$ generally corresponds to the isophote's semi-major axis ($R_{\rm maj}$). Often times, however, the major axis profile is mapped onto the so-called `equivalent', or geometric mean axis, $R_{\rm eq}$, through:

\begin{equation}
R_{\rm eq} = R_{\rm maj}\sqrt{1 - \epsilon(R_{\rm maj})},
\label{equ:eq}
\end{equation}

where $\epsilon(R_{\rm maj})$ is the isophote ellipticity, defined as 1 minus the axis ratio. This mapping converts the isophote into the equivalent circle that conserves the original surface area of the isophote (see the Appendix of \citealt{Ciambur2015} for a derivation). The equivalent axis profile is thus circularly symmetric, and decomposing it allows for an analytical computation of the total magnitude of components directly from their parameters (e.g., \citealt{Graham&Driver2005} for S\'ersic parameters).

The user has a choice between modelling the profile along $R_{\rm maj}$ (the default) or $R_{\rm eq}$. For the latter option, provided that the input data contains ellipticity information, \prof~generates the equivalent axis profile internally and outputs the total magnitudes of components after the decomposition. If the input data is a two-column table, it is assumed that the $R$ column is already the axis chosen by the user. In order to avoid convolution issues arising from non-uniform radial sampling (necessarily arising from Equation \ref{equ:eq}, unless $\epsilon$ and the step in $R_{\rm maj}$ are both constant; also, when the profile sampling step is logarithmic), \prof~first linearly interpolates the SBP on a uniformly spaced radial axis.

\subsection{The Point Spread Function}

The ability of telescopes to resolve a point-source is dictated by a number of factors, including their diffraction limit (due to the fact that they have a finite aperture), the detector spatial resolution (pixel size) and, for ground-based instruments, the distortion of wavefronts caused by turbulent mixing in the atmosphere, an effect known as `seeing'. All of these effects blur astronomical images, spreading the light at every point in a way characteristic to each instrument. In an ideal image, a point source's profile is a delta function. In a real image, however, the functional form is called the instrumental {\it point spread function} (PSF). In order to reconstruct the true distribution of light in an image it is essential to know the PSF (at every point\footnote{The PSF changes with position on the focal plane.}).

The most basic aproximation of a PSF is a Gaussian functional form with the single parameter FWHM (or dispersion $\sigma$, the two being related by FWHM = $2\sigma\sqrt{2\,{\rm ln}2}$). This form, however, underestimates the flux in the `wings' of the PSF, which can bias decomposition parameters (\citealt{Trujillo+2001} found the effect to range between 10--30\% for S\'ersic parameters). 

A more realistic aproximation which $is$ capable of modelling PSF wings is the Moffat profile (\citealt{Moffat1969}), given by:

\begin{equation}
\label{equ:moffat}
I(R) = I_0 \,\left[1+\left( \frac{R}{\alpha}\right)^2\right]^{-\beta}
\end{equation}

where $\alpha$ is a characteristic width related to the FWHM by the identity FWHM  = $2\alpha\sqrt{2^{1/\beta}-1}$, and $\beta$ controls the amount of light in the `wings' of the profile compared to the centre (redistributing the light of the central peak into wings mimics the effect of spreading light in Airy rings). Figure \ref{fig:psf} shows Moffat functions of the same FWHM but different values of $\beta$, as well as the limiting case where $\beta \rightarrow \infty$, which corresponds to a Gaussian (\citealt{Trujillo+2001}).

\begin{figure}
\includegraphics[width=1.\columnwidth]{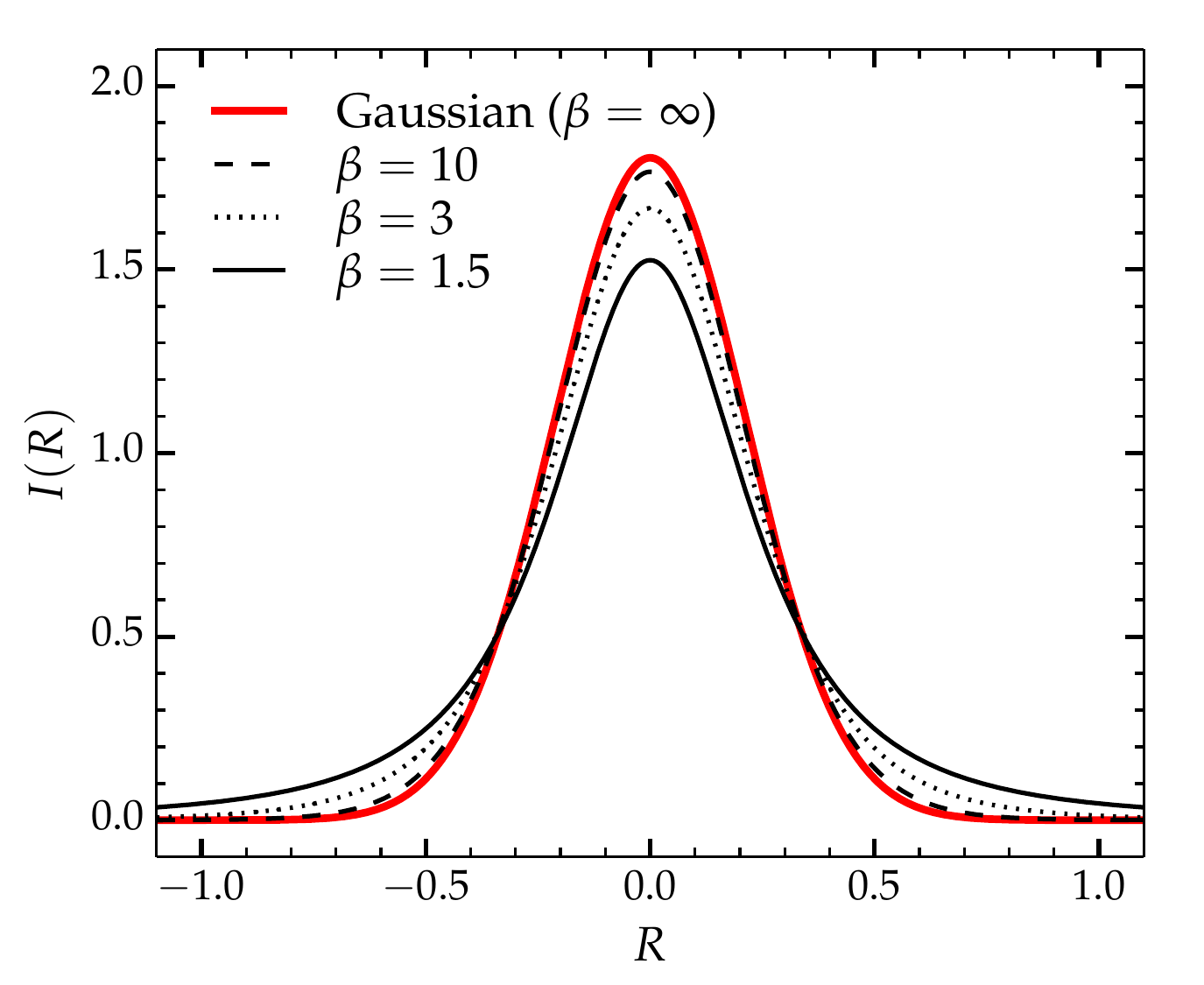}
\caption{The Moffat function (black curves) accounts for seeing effects (e.g., Airy rings) by transferring flux from the peak of the PSF into its wings. This is controlled by the $\beta$ parameter and, for large values of $\beta$ the Moffat approaches a Gaussian (red curve). All curves plotted here have a FWHM of 0.5, and axes are in arbitrary units.}
\label{fig:psf}
\end{figure}

In practice, when characterising the PSF the usual norm is to fit either a Gaussian or a Moffat profile on bright, unsaturated stars in the image with e.g., the IRAF task {\sc Imexamine}. This task directly provides the FWHM for the former and (FWHM; $\beta$) for the latter case. 

In \prof~the user has a choice of either Gaussian, Moffat or data vector PSF. The first two require the parameters specific to each function, from which \prof~generates the PSF internally when needed. The third option requires a table of values, $R$ and $I(R)$, in the form of a text file provided by the user\footnote{This can be obtained by e.g., extracting the light profile of a bright star in the image.}. The radial extent of the data vector PSF is required by \prof~to at least match or exceed that of the galaxy profile. As I will show in Section \ref{sec:p52} with an example, this feature is very useful when the analytical functions above do not provide a sufficiently exact description of the PSF.
 
\section{THE MODEL}\label{sec:model}

\prof~can employ several analytical functions to model the radial light profiles of a galaxy's constituent components. The user is free to add an indefinite number of components to the model, and each component (function) can have its parameters freely varying or fixed to given values during the fit.

In the remainder I provide a description of each function available in \prof~in the context of the photometric component(s) which it is intended to model.

\subsection{Ellipticals and Galaxy Bulges}

\subsubsection{The S\'ersic Model}

There have been many attempts in the past to analytically describe the SBPs of elliptical galaxies, including deVaucouleurs' $R^{1/4}$ `law' (\citealt{deVaucouleurs1948,devaucouleurs1953}), the King profile (\citealt{King1962,King1966}) etc. (see the review by \citealt{Graham2013}). At present it is generally agreed that the most robust function for this purpose is given by the \cite{Sersic1963} $R^{1/n}$ model (\citealt{Caon+1993}, \citealt{D'Onofrio+1994}). 

While the S\'ersic function in itself does not contain any physical meaning, it is remarkably flexible and can accurately capture the light profiles of a broad range of spheroid components, from the small bulges of late-type spiral galaxies to the highly concentrated light profiles of bright elliptical galaxies. Additionally (as will be discussed in the following sections) the S\'ersic profile can also model discs and bars.

The S\'ersic profile is parameterised by 3 quantities: the radius enclosing half of the light, $R_e$, the intensity at this radius, $I_e = I(R_e)$, and the concentration, or S\'ersic index, $n$. It takes the form:

\begin{figure}
\includegraphics[width=1.\columnwidth]{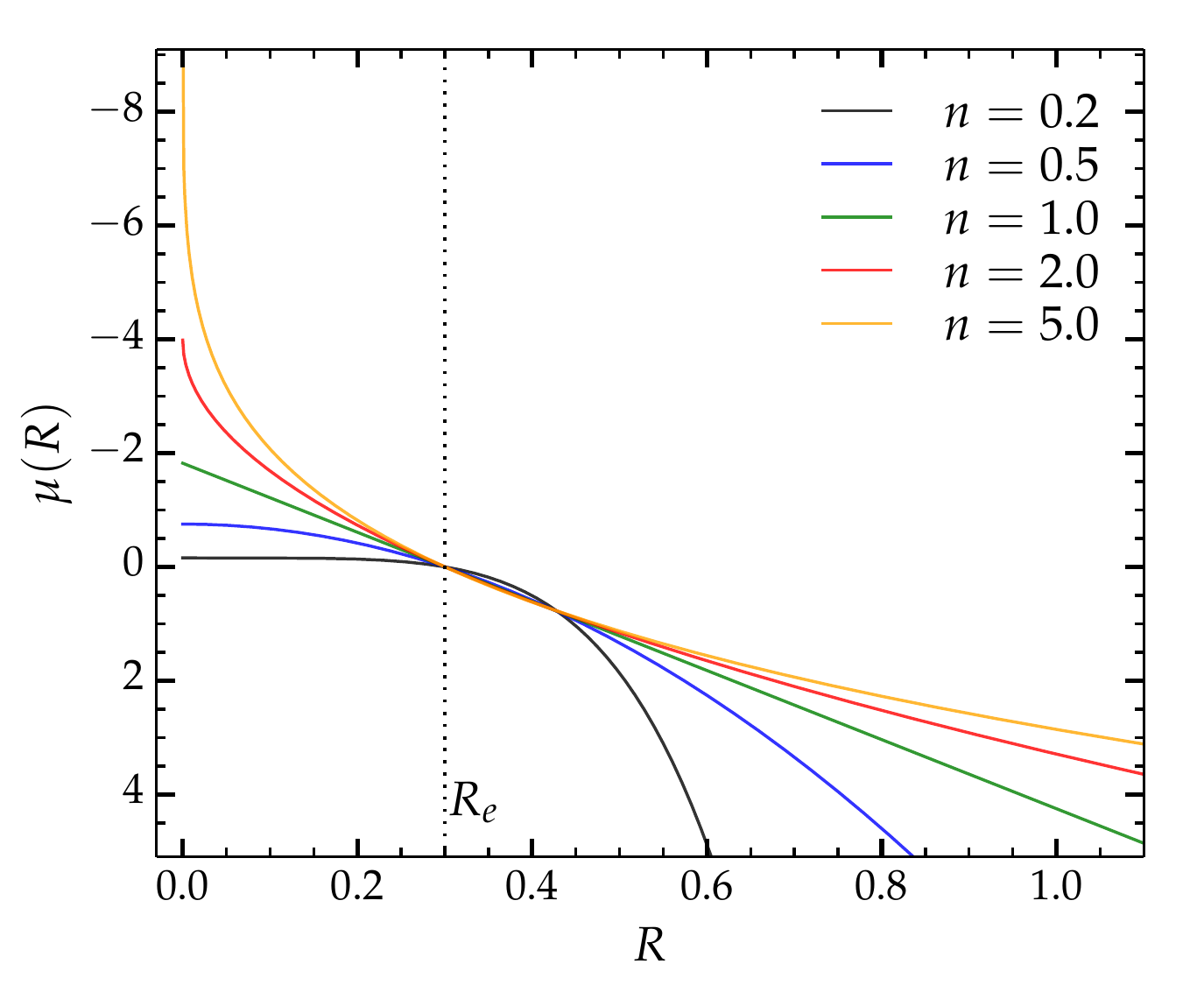}[h]
\caption{The S\'ersic profile for five values of the S\'ersic index $n$. The axes are in arbitrary units, but all profiles have the same values of $\mu_e$ and $R_e$. The half-light radius, $R_e$, is indicated by the vertical dotted line.}
\label{fig:sersic}
\end{figure}

\begin{equation}
I(R) = I_e {\rm exp} \left\{ -b_n \left[ \left( \frac{R}{R_e} \right)^{\frac{1}{n}} -1 \right] \right\}, 
\label{equ:sersic}
\end{equation}

\noindent where $b_n$ depends on $n$ and is obtained by solving:

\begin{equation}
\Gamma (2n) = 2\gamma (2n, b_n), 
\label{equ:Gamma}
\end{equation}

\noindent where $\Gamma$ is the (complete) gamma function and $\gamma$ the incomplete gamma function, given by the integral:

\begin{equation}
\gamma (2n,x) = \int_0^{x} {\rm e}^{-t} t^{2n -1} {\rm d}t
\label{equ:gamma}
\end{equation}

The reader will also find a review of the S\'ersic model, useful equations pertaining to it, as well as early references, in \cite{Graham&Driver2005}.

\subsubsection{The core-S\'ersic Model}

The most luminous early-type galaxies display `cored' central profiles, thought to be the result of black hole binary systems kicking out stars through 3--body interactions (\citealt{Begelman+1980}), thus causing a deficit of light in the centre (\citealt{King1978}, \citealt{Dullo&Graham2012, Dullo&Graham2014} and references therein). An ideal functional form which describes these types of objects is the 6-parameter core-S\'ersic model (\citealt{GrahamEA2003}), given by:

\begin{equation}
I(R) = I' \left[ 1 + \left( \frac{R_b}{R} \right)^{\alpha} \right]^{\frac{\gamma}{\alpha}}  {\rm exp} \left\{ -b_n \left[ \frac{R^{\alpha} + R_b^{\alpha}}{R_e^{\alpha}} \right]^{\frac{1}{\alpha n}} \right\}.
\label{equ:core-sersic}
\end{equation}

The core-S\'ersic function takes the form of a power law in the core region, which then transitions into a S\'ersic form outside the core region (Figure \ref{fig:core-sersic}). It is parameterised by the break (transition) radius $R_b$ and half-light radius $R_e$, the inner profile slope $\gamma$, the smoothness of the transition, controlled by $\alpha$, the S\'ersic index $n$ and a normalisation, or scale intensity $I'$, which is related to the intensity at the break radius through eq. 6 in \cite{GrahamEA2003}, that is:

\begin{equation}
\label{equ:I_primed}
I' = I_b\;2^{-\gamma/\alpha}\;{\rm exp} \left[ b_n \left( \frac{2^{1/\alpha} R_b}{R_e} \right)^{1/n} \right]
\end{equation}

\begin{figure}[h]
\includegraphics[width=1.\columnwidth]{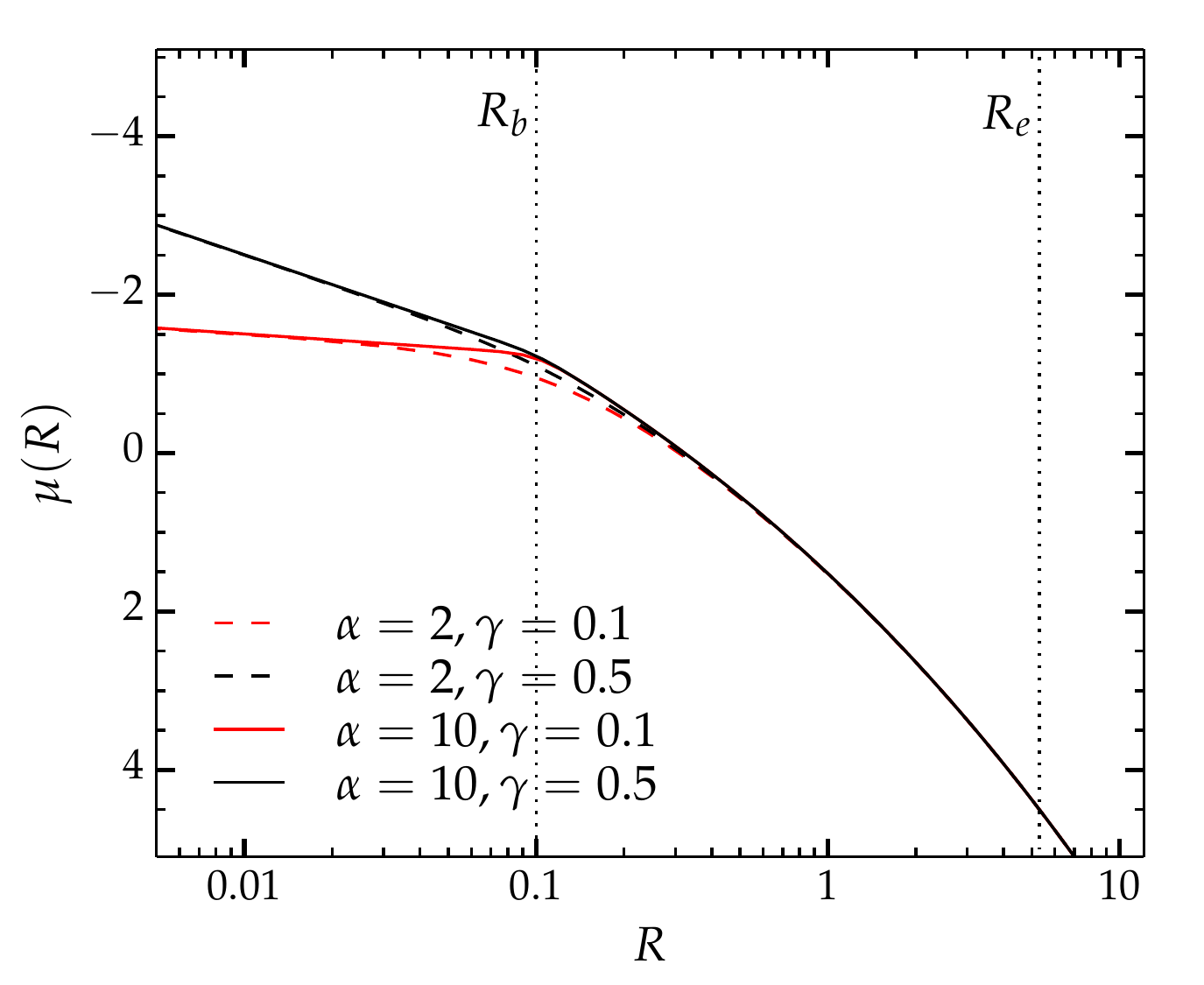}
\caption{The core-S\'ersic profile: varying the inner slope $\gamma$ (red curves) and the break sharpness $\alpha$ (black curves). The break radius and effective radius are indicated through the vertical dotted lines, and are marked as $R_b$ and $R_e$, respectively. Axes are in arbitrary units.}
\label{fig:core-sersic}
\end{figure}

I note in passing that the core-S\'ersic model has also been recently implemented in 2D fitting codes, specifically in the {\sc Galfit} add-on code {\sc Galfit-Corsair} (\citealt{Bonfini2014}), and in {\sc Imfit} (\citealt{Erwin2015}).

\subsection{Discs}

\subsubsection{The Exponential Model}

The radial decline of light in flat galaxy discs has long been known to be approximately exponential (\citealt{Patterson1940}, \citealt{deVaucouleurs1957}, \citealt{Freeman1970}). For disc components, \prof~can employ the two-parameter exponential model:

\begin{equation}
I(R) = I_0\, {\rm exp}\left(-\frac{R}{h}\right),
\label{equ:exponential}
\end{equation}

\noindent where $I_0$ is the intensity at $R=0$ and $h$ is the exponential scale length, which corresponds to the length in which the light diminishes by a factor of $e$, i.e., $I(h)=I_0/e$.

\subsubsection{The Broken Exponential Model}

While the light profiles of galaxy discs are commonly assumed to be characterised by a single exponential for their whole extent (galaxies for which this is true are said to have `Type I' profiles in the classification of \citealt{Erwin+2008}), as many as 90\% (\citealt{Pohlen&Trujillo2006}) of disc galaxies display a `break' in their light profiles, thypically at a few scale lengths from the photocentre (\citealt{vanderKruit1987}, \citealt{Pohlen+2004}). More specifically, this is an abrupt change in their exponential scale length (Figure \ref{fig:truncated-disc}). This phenomenon is referred to as a {\it truncation}, or `Type II' profile, if the scale length decreases (the light fall-off becomes steeper) and an {\it anti-truncation}, or `Type III' profile (\citealt{Erwin+2005}), if the scale-length becomes longer (the fall-off becomes shallower). \prof~provides a broken exponential function to fit these types of profiles (Equation \ref{equ:trunc}), characterised by four free parameters: the central flux $I_0$, the break radius $R_b$, and the inner and outer scale lengths, $h_1$ and $h_2$, respectively.

\begin{equation}
I(R) = 
	\begin{cases}
	I_0\, {\rm exp} \left(-R/h_1\right), & R \leqslant R_{b}\\
	I_{b}\, {\rm exp} \left[-(R-R_{b})/h_2\right], & R > R_{b} \:,
	\end{cases}
\label{equ:trunc}
\end{equation}

\noindent where $I_b$ is the brightness at the break radius, and is not a free parameter since $I_b = I(R_b)$.\\ 

\begin{figure}[h]
\includegraphics[width=1.\columnwidth]{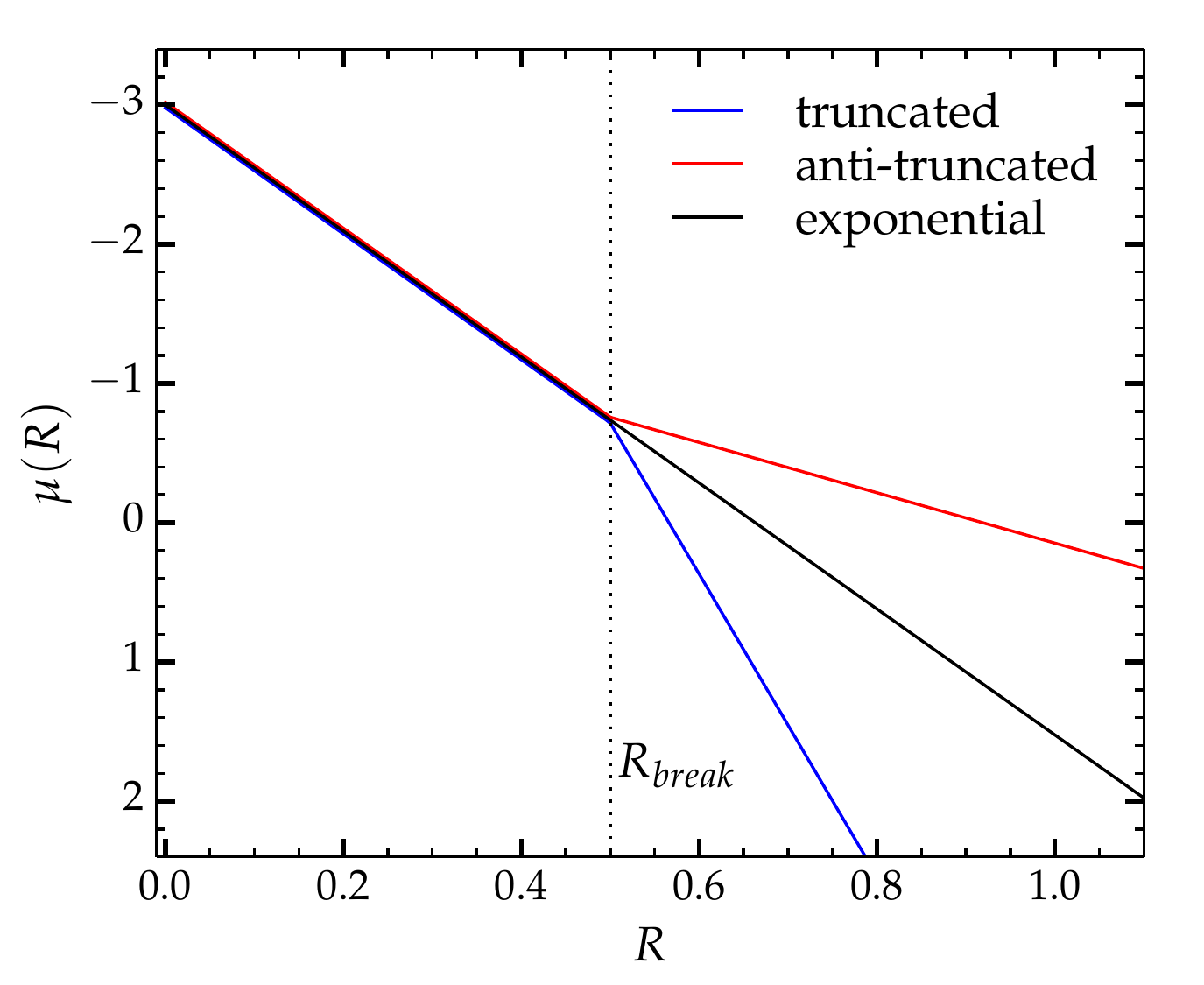}
\caption{The broken exponential profile: a Type II, or truncated, profile ($h2 < h1$, blue), a Type III, or anti-truncated profile ($h2 > h1$, red). The black curve is a single exponential (Type I) profile, for comparison. Axes are in arbitrary units.}
\label{fig:truncated-disc}
\end{figure}

\subsubsection{The Edge-on Disc Model}

When disc galaxies are viewed in close to edge--on orientation, the disc SBP exhibits a gradually shallower slope towards the centre (\citealt{Kruit&Searle1981}; \citealt{Pastrav+2013}). In this regime, \prof~can use two special cases of the inclined disc profile of \cite{Kruit&Searle1981}, which is defined in 2D as a function of major axis $R$ and minor axis $Z$, as:

\begin{equation}
\label{equ:inclined-disc}
I(R,Z) = I_0\, \left( \frac{R}{h_r} \right) \,K_1\left( \frac{R}{h_r} \right)\, {\rm sech}^2\left( \frac{Z}{h_{z}} \right) ,
\end{equation}

\noindent where $I_0$ is the central intensity, $h_r$ is the scale length in the plane of the disc (major axis), $h_z$ is the scale length in the vertical (minor axis) direction ($\perp$ to the disc plane), and $K_1$ is the modified Bessel function of the second kind.

In the limiting case of $Z=0$, Equation \ref{equ:inclined-disc} reduces to the major axis form:

\begin{equation}
\label{equ:inclined-disc-major}
 I(R_{\rm maj}) = I_0\, \left( \frac{R_{\rm maj}}{h_r} \right) \,K_1\left( \frac{R_{\rm maj}}{h_r} \right) ,
\end{equation}

Similarly, along the minor axis ($R=0$), Equation \ref{equ:inclined-disc} reduces to:

\begin{equation}
\label{equ:inclined-disc-minor}
I(R_{\rm min}) = I_0\,  {\rm sech}^2\left( \frac{R_{\rm min}}{h_z} \right)
\end{equation}

\begin{figure}[h]
\includegraphics[width=1.\columnwidth]{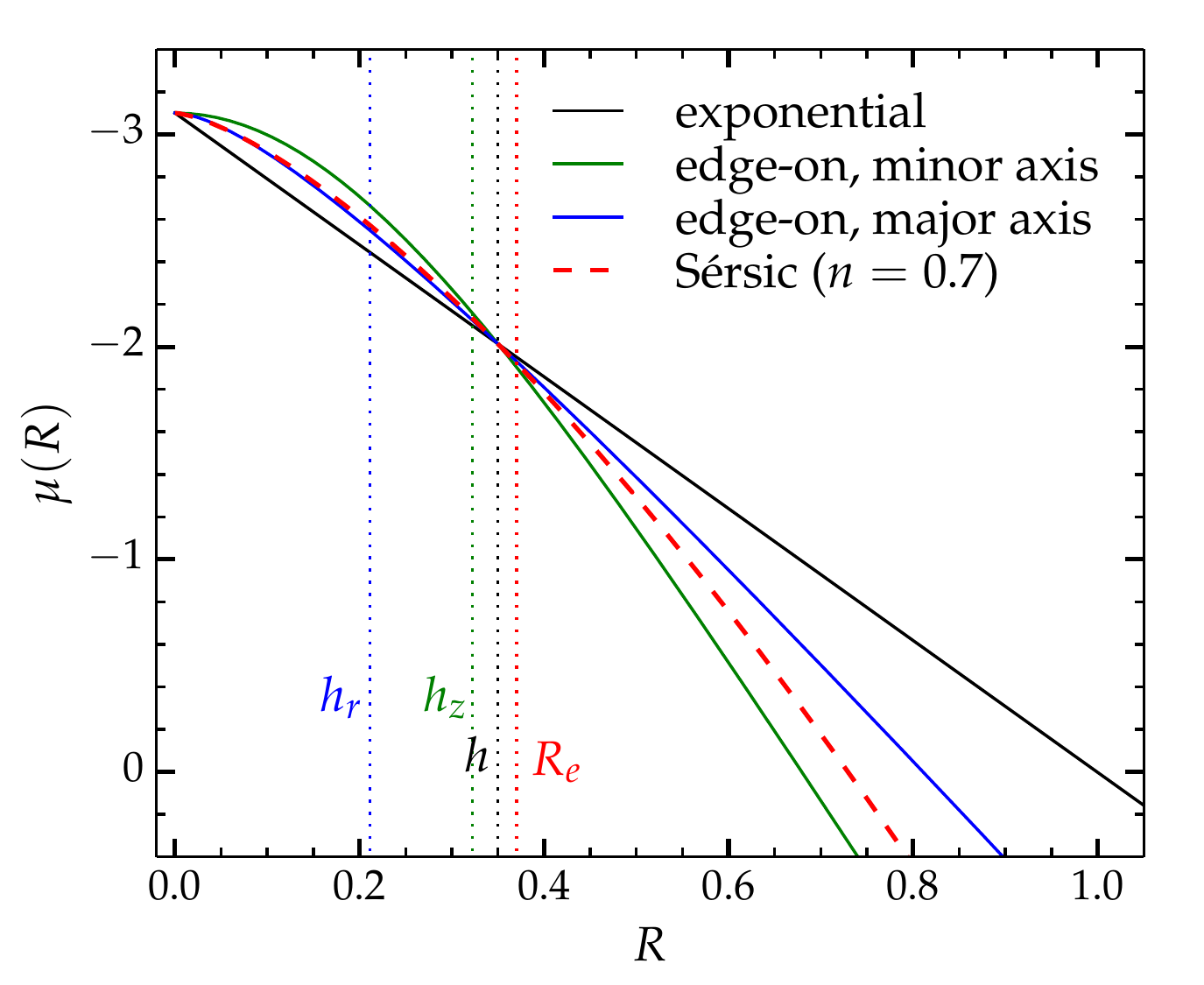}
\caption{The possible disc components: Exponential (black, solid), edge-on disc along major-axis (Equation \ref{equ:inclined-disc-major}; blue, solid), edge-on disc along the minor-axis (Equation \ref{equ:inclined-disc-minor}; green, solid) and S\'ersic ($n <1$; red, dashed). All profiles have the same central surface brightness ($\mu_0$) and the same $e$--folding radius, equal to $h$, the exponential scale length. The vertical dotted lines mark each profile's characteristic scale length (keeping the colour scheme).  As in Figure \ref{fig:sersic}, the axes are in arbitrary units.}
\label{fig:inclined-disc}
\end{figure}

\prof~can employ Equations \ref{equ:inclined-disc-major} and \ref{equ:inclined-disc-minor} to fit edge-on discs along the major and minor axes, respectively. Note that neither $h_r$ nor $h_z$ are equal to the $exponential$ scale length $h$. Their corresponding profiles do not decrease in intensity by a factor of $e$ at every $h_r$ or $h_z$. One can readily discern this visually in Figure \ref{fig:inclined-disc} (solid curves): the curvature of these profiles towards $R \rightarrow 0$ implies that they cannot be characterised by a single value for the $e$--folding radius, as exponential profiles are. Rather, towards the centre (where the slopes are shallower), the $e$--folding is large, whereas at increasing radii, the $e$--folding decreases and asymptotes to a constant value at high radii, where both profiles approach an exponential form. Again, the scale lengths in Equations \ref{equ:inclined-disc-major} and \ref{equ:inclined-disc-minor} are not these asymptotic values (see \citealt{Kruit&Searle1981} for more details on their definition). The major-axis profile has an $e$--folding length (from $R = 0$) equal to 1.658$h_r$, whereas the minor-axis profile has an $e$--folding length equal to 1.085$h_z$\footnote{This factor is given by $1.085 = {\rm arccosh}(\sqrt{e})$, not  to be confused with the factor $1.086 = 2.5/{\rm ln}(10)$ in Eq. 17 from \citealt{Graham&Driver2005}, which relates the central surface brightness of an $exponential$ to the value at $R=h$ by $\mu_0 = \mu(h) - 1.086$.}. This is illustrated in Figure \ref{fig:inclined-disc}.

\subsubsection{The S\'ersic Model for Discs}

Finally, as mentioned before, the S\'ersic function can also successfully model discs. The $n=1$ S\'ersic function is identical to an exponential and can be used to model Type I profiles, while inclined (edge-on) discs, or those with dusty centres, may be fitted with $n<1$ S\'ersic functions (typically in the range $n \sim 0.7 - 0.9$; see Figure \ref{fig:inclined-disc}). In this case, one can still recover its $e$--folding length $h_S$ and central surface brightness $\mu_0$ from:

\begin{equation}
\label{equ:scale-length}
h_S = \frac{R_e}{(b_n)^n}\:\left(= \frac{R_e}{1.67835} \:\:{\rm for}\:\: n=1 \right) ,
\end{equation}

\noindent and

\begin{equation}
\label{equ:central-sbp}
\mu_0 = \mu_e - \frac{2.5}{{\rm ln}(10)}b_n\:\left( = \mu_e - 1.82224 \:\:{\rm for}\:\: n=1 \right) ,
\end{equation}

\noindent though, as before, if $n \neq 1$, the $e$--folding radius is not unique along the whole profile and is again highest towards the centre and lower at high-$R$ (unlike the edge-on disc model, the $n < 1$ S\'ersic function does not asymptote to an exponential, but has an ever steeper slope with increasing $R$; see Figure \ref{fig:sersic}).

\subsection{Bars}

Bars are common in disc galaxies (the fraction is $\sim 2/3$ in the near-infrared; \citealt{Eskridge2000}, \citealt{Menendez-Delmestre+2007}; see also \citealt{Laurikainen+2009} and \citealt{Nair&Abraham2010}) and display a characteristic flat profile which ends with a relatively sharp drop-off. This shows up in SBPs as a `shelf'-like or `shoulder'-like feature, usually (but not necessarily) between the inner spheroid component and the outer disc. Bars are often modelled with the four-parameter Ferrers profile (\citealt{Ferrers1877}), expressed as:

\begin{equation}
\label{equ:ferrers}
I(R) = I_0\,  \left[ 1- \left(\frac{R}{R_{\rm end}} \right)^{2-\beta} \right]^{\alpha},
\end{equation}

\begin{figure}[h]
\includegraphics[width=1.\columnwidth]{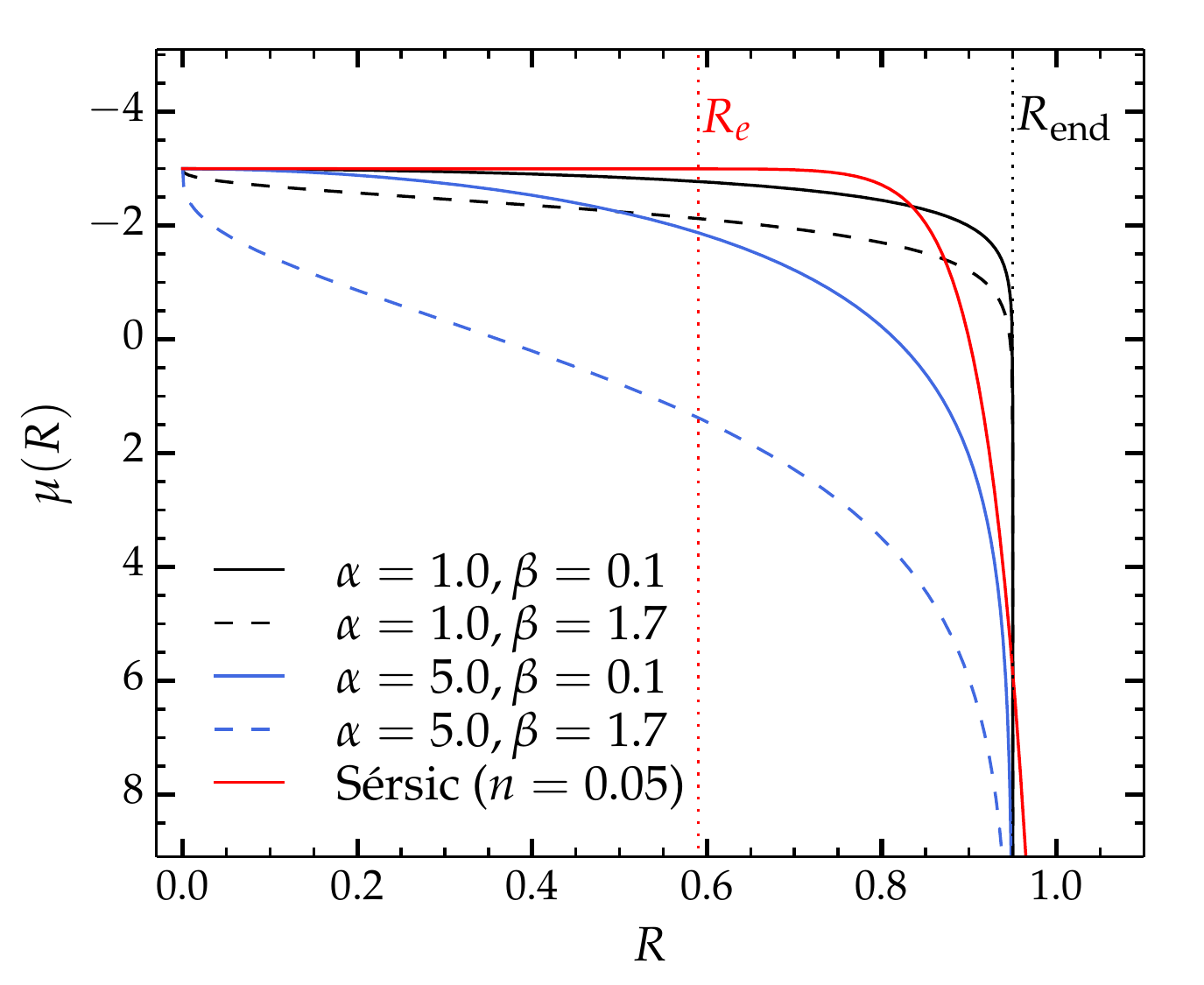}
\caption{Possible bar components: The black and blue curves are all Ferrers profiles (Equation \ref{equ:ferrers}) with the same central surface brightness ($\mu_0$) and end radius ($R_{\rm end}$), but different permutations of the $\alpha$ and $\beta$ parameters, such that curves of the same colour illustrate the effect of changing $\beta$, while curves of the same style (i.e., solid vs. dashed) illustrate the effect of changing $\alpha$.}
\label{fig:bar}
\end{figure}

\noindent where $I_0$ is the central brightness, $R_{\rm end}$ is the cut-off radius, and $\alpha$ and $\beta$ control the inner slope and break sharpness (see Figure \ref{fig:bar}). Note that $\beta > 0$ causes a cusp in the central parts of the profile. As there is no observational evidence that the profiles of bars display such a cusp, it is recommented that $\beta$ be set to 0. 

Another function which can be used to model bars is a low-$n$ S\'ersic function (typically in the range $n \lesssim 0.1-0.5$, the drop-off being sharper the lower $n$ is; see Figure \ref{fig:bar}, red curve). 

\begin{table*}
\centering
\caption{An index of the functions available in \prof.}
\begin{tabular}{l l l}
\hline
Function & Parameters & Used for \\
\hline
\hline \\

S\'ersic & 3 & ellipticals, bulges, bars, edge-on discs\\
core-S\'ersic & 6 & cored ellipticals\\
exponential & 2 & discs\\
broken exponential & 4 & (anti)truncated discs\\
edge-on disc model & 2 & edge-on discs\\
Ferrers & 4 & bars\\
Gaussian & 3 (1/0) & rings, spiral arms, (point source/ PSF)\\
Moffat & 1/0 & point source/ PSF\\
data vector PSF & 1/0 & point source/PSF\\
\hline
\end{tabular}
\label{tab:functions}
\end{table*}

\subsection{Rings and Spiral Arms}

The presence of spiral arms in a disc can cause deviations from an exponential profile in the form of `bumps' (\citealt{Erwin+2005}). A stellar ring also causes a `bump' in the profile. These features are modelled in \prof~ via the three-component Gaussian profile, given by:

\begin{equation}
\label{equ:gaussian}
I(R) = I(R_r) \, {\rm exp}\left[ -\frac{(R-R_r)^2}{2\sigma^2}\right],
\end{equation}

\noindent where $R_r$ is the radius of the bump, $I(R_r)$ is its peak intensity and $\sigma$ its width (dispersion).

\section{FITTING THE DATA}\label{sec:fitting}

Once the required input information is provided, the user must make a choice of components that are to make up the model. Before this (or at any point after having provided the input information), they can visualise the galaxy light profile and, if it is based on isophote fits with {\sc Ellipse} or {\sc Isofit}, additional information such as the ellipticity $\epsilon$, position angle (PA) or $B_4$ ($4^{\rm th}$ cosine harmonic amplitude) profiles can be displayed. This helps the user decide which components to use and make an educated guess for the initial values of their parameters. The default setting is that all model parameters are free, but the user has a choice to hold one or more of the parameters fixed to specific values during the fitting process. 

An additional user requirement is a value for a `global' ellipticity of the central profile ($\epsilon_c$), i.e. the part dominated by the PSF. $\epsilon_c$ is important for the convolution process because models with different ellipticities yield slightly different convolved SBPs, as will become clear in Section 4.1. Often, however, the galaxy being modelled consists of a superposition of components, each with its own ellipticity, so a single value for $\epsilon$, naturally, does not apply to the entire model. However, $\epsilon_c$ is only relevant for the part of the model affected by the PSF, and should roughly correspond to the ellipticity of that component which dominates the light in the central few PSF FWHM. 

The user can estimate $\epsilon_c$ as a luminosity-weighted average of the galaxy's ellipticity profile at a radius of $\sim$ 2--3 PSF FWHM. The values interior to this should be avioded because here the isophotes are circularised by the PSF, and do not reflect the component's ellipticity. If $\epsilon_c$ is not provided, the default value is set to zero, which corresponds to a circularly symmetric model.

When all the desired components are included, \prof~can begin the search for the best-fitting solution through an iterative minimization process, which begins with the parameter guess-values set by the user. Each iteration (corresponding to a specific location in the parameter space) consists of generating a model (which consists of the sum of components), convolving it with the PSF, and comparing the result with the data. 

\subsection{PSF Convolution}

The convolution of the model with the PSF is performed in two dimensions (2D) due to two important aspects. 

First, the axis ratio (or ellipticity) of a component affects the way its light distribution is blurred by seeing effects (see \citealt{Trujillo+2001b, Trujillo+2001}). For a circularly symmetric PSF, if the component too is circularly symmetric then the light from a point located at a distance $R$ from the centre is scattered the same way as any other point at the same radius $R$. However, if the component is elliptical, then the light at fixed distance from the centre is scattered more efficiently by points lying on the major axis than by the points in other azimuthal directions.  In this case, the convolved major axis profile is systematically lower than in the circular case. This effect is illustrated in Figure \ref{fig:Ellipticity_effect}, for three S\'ersic profiles, each with three different axis ratios.

\begin{figure*}
\includegraphics[width=0.32\textwidth]{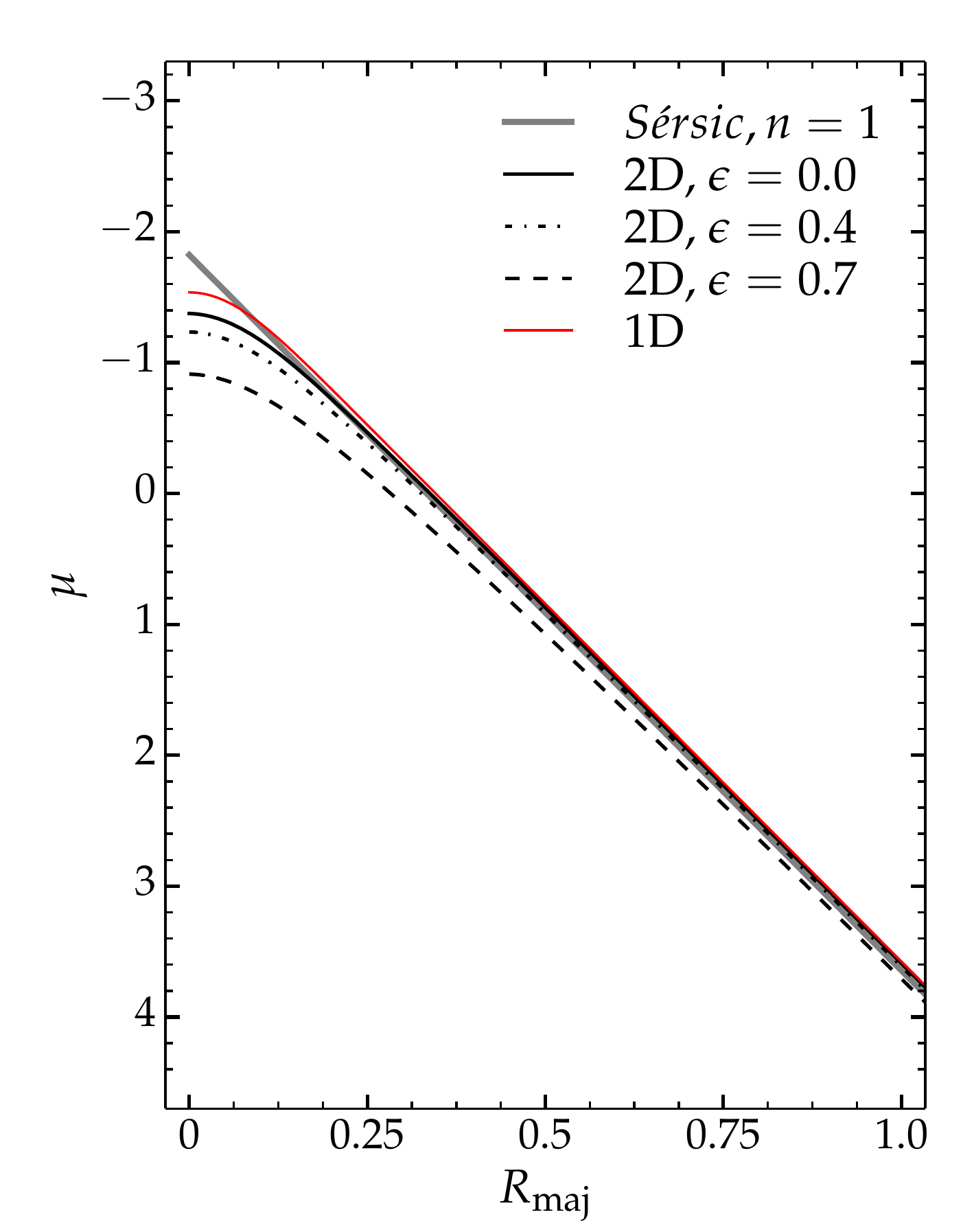} 
\includegraphics[width=0.32\textwidth]{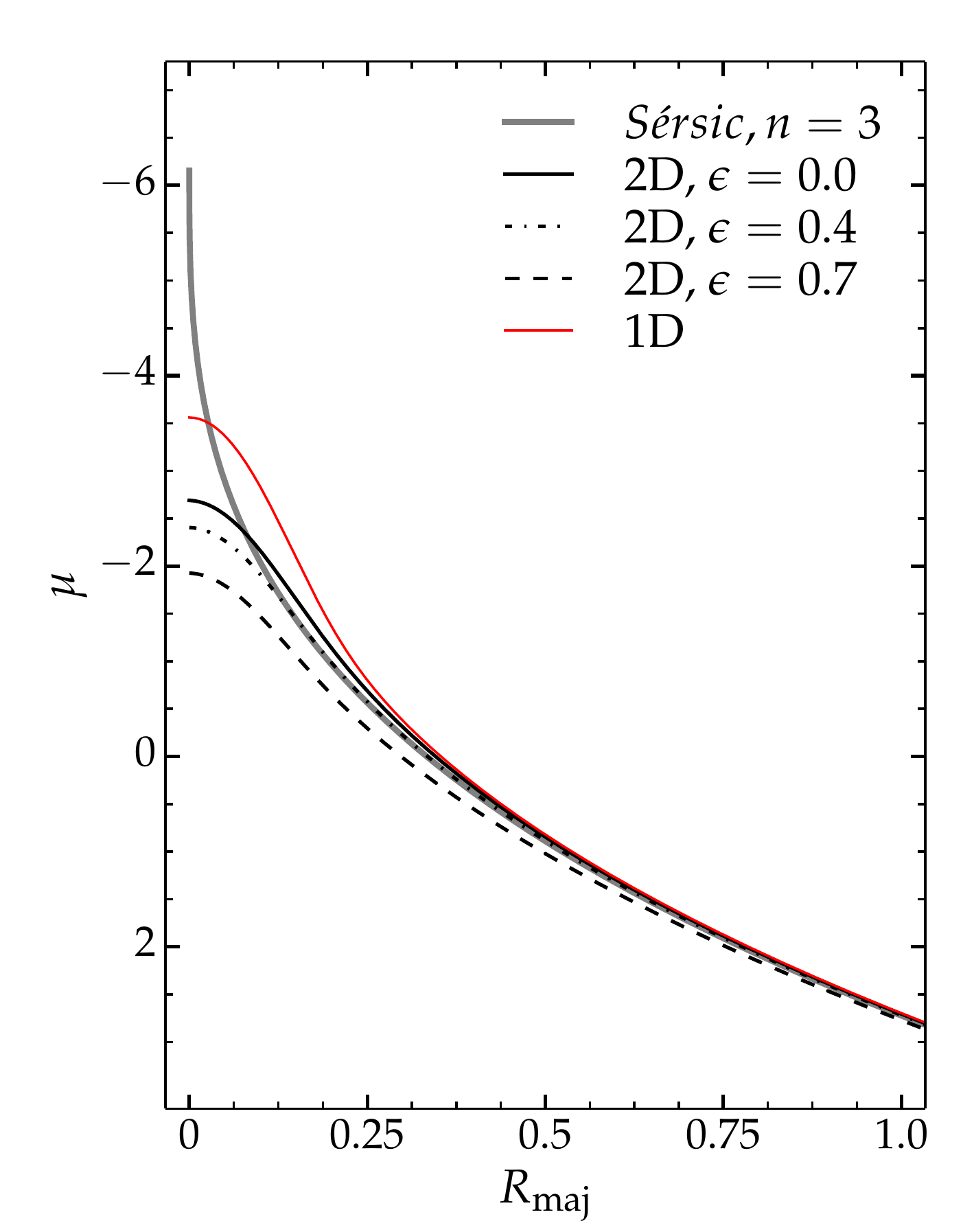} 
\includegraphics[width=0.32\textwidth]{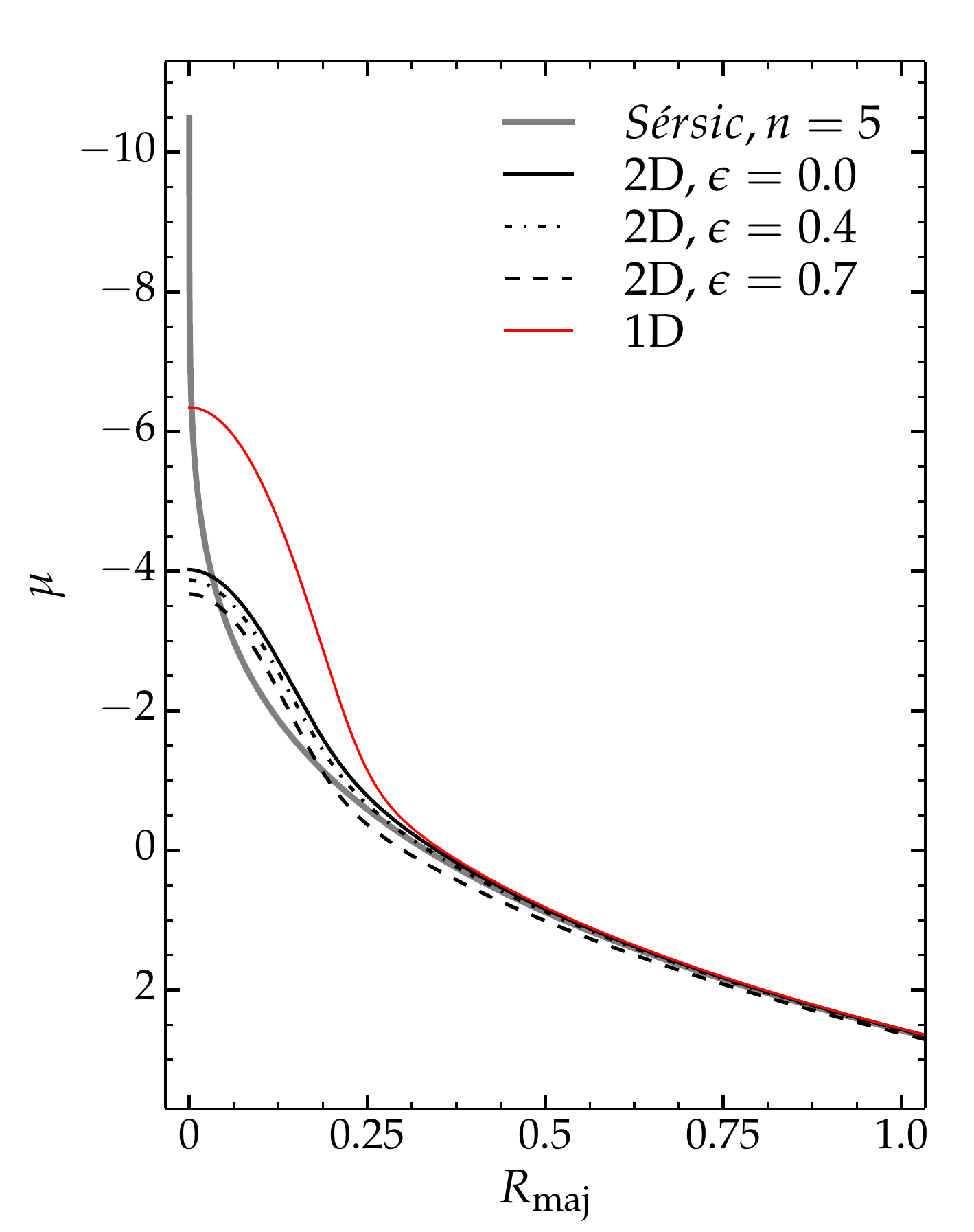}
\caption{The effect of a component's ellipticity on the PSF convolution, for three different S\'ersic functions with the same $\mu_e$ and $R_e$ but different $n$. In each panel, the thick grey curve represents the unconvolved S\'ersic profile, the thick black lines (solid, dot-dashed and dashed) represent the profile convolved in 2D with a circular Gaussian PSF, assuming different ellipticities ($\epsilon$) for the S\'ersic model, while the thin red curve represents the profile (incorrectly) convolved in 1D with the same PSF profile. Convolving the grey curve in 1D is inappropriate because, while it does conserve the area under grey curve (Equation \ref{equ:int1}), it does not conserve the total flux (Equation \ref{equ:int2}), and therefore does not model the effect of seeing. The Gaussian FWHM was chosen to be large (0.167 = 0.5$R_e$, in the arbitrary units of the $x$-axis), for clarity.}
\label{fig:Ellipticity_effect}
\end{figure*}

Second, the PSF convolution should not be performed in one dimension (1D), i.e., by convolving the SBP curve with the PSF profile. This is because each data point in the SBP represents a (local) $surface$ brightness at a particular distance from the centre ($R$) and along a particular direction ($\theta$), i.e., $I \equiv I(R,\theta)$. The SBP itself represents the radial distribution of light {\it for a particular choice of $\theta$} (e.g., major axis: $\theta = 0$), and is analogous to (but more accurate than) a 1D cut from the image\footnote{Provided that the isophote PAs are constant with radius.}. As such, the area under the curve,

\begin{equation}
\label{equ:int1}
\int_0^{\infty} I(R,\theta){\rm d}R,  
 \end{equation}
 does not correspond to the total flux in the image, which requires an additional integration in the azimuthal direction: 
 
 \begin{equation}
 \label{equ:int2}
\int_0^{2\pi} \int_0^{\infty} R\, I(R,\theta) \: {\rm d}R\, {\rm d}\theta.
 \end{equation}
 A 1D convolution conserves the area under the SBP curve (Equation \ref{equ:int1}) but not the total flux in the image (Equation \ref{equ:int2}), and is therefore inappropriate to reproduce seeing effects in images, which conserve total flux. The difference between the two types of convolution (1D vs 2D) is also illustrated in Figure \ref{fig:Ellipticity_effect}.

The convolution is performed in several steps. If the fitting axis is the major axis, then an elliptical 2D light distribution is generated, with a global ellipticity provided by the user and the same profile along the major axis as that of the model SBP. If the fitting is performed along the equivalent axis, the convolution is performed as above except that $\epsilon_c$ is set to 0, since the equivalent axis is circularly symmetric by construction. 

In the next step, a circularly symmetric 2D PSF is generated, on the same array as the model. This can be based on either the parameters of Gaussian or Moffat forms, or the user-provided data vector PSF (depending on the PSF choice). The model array and PSF array are then convolved using the FFT (Fast Fourier Transform) method, and finally, the resulting distribution's major axis profile is obtained, which represents the desired convolved model SBP. It is this quantity that is compared, at each iteration, with the data.

\subsection{Minimisation and Solution}

While most 2D decomposition codes perform signal-to-noise (S/N) -weighted minimisations in intensity units, \prof~uses an un-weighted least-squares method in units of surface brightness. Because noise in galaxy images is dominated by Poisson noise, galaxies have the highest S/N at the centre, where they are (usually) brightest. However, the central regions can often be affected by dust, active galactic nuclei, nuclear discs, or PSF uncertainty (the latter being particularly important for highly concentrated galaxies), so placing most of the weight on the central data points can substantially bias the fit for the entire radial range if all of these issues are not dealt with (see similar arguments in \citealt{Graham+2016}). In an un-weighted scheme, all data points along the SBP contribute equally to constraining the model and thus, even when the central data is biased, the model is still well constrained by the outer data points. The caveat, however, is that one must ensure that the image sky background is accurately measured and subtracted. Otherwise, the outer data points introduce a bias in the model (e.g., if the outer data corresponds to an exponential disc, inaccurate background subtraction would lead to the wrong scale length, which affects the entire radial range because a disc component runs all the way to the centre.)

The minimisation is performed with the {\sc Python} package {\sc lmfit}\footnote{\url{https://pypi.python.org/pypi/lmfit/}}, and the method is a least-squares Levenberg-Marquardt algorithm (\citealt{Marquardt1963}). The quantity being minimized is: 

\begin{equation}
\label{equ:deltarms}
\Delta_{\rm rms} = \sqrt{ \sum_i (\mu_{{\rm data},i} - \mu_{{\rm model},i })} , 
\end{equation}

\noindent where $i$ is the radial bin, $\mu_{\rm data}$ is the data surface brightness profile and $\mu_{\rm model}$ is the model at one iteration.

When the solution is reached, the result is displayed and a logfile is generated. The logfile contains a time-stamp of the fit, all the input information and settings, and a raw fit report with all parameters and their correlations. A more in-depth report follows, which lists each component's best-fit parameter values and, if the decomposition was performed on the equivalent axis, their total magnitude.

The quantity $\Delta_{\rm rms}$ quantifies the global quality of the fit,  but a more reliable proxy of the solution's accuracy, in detail, is the residual profile: $\Delta\mu(R) = \mu_{\rm data}(R) - \mu_{\rm model}(R)$. A good fit is characterised by a flat $\Delta\mu$ profile which scatters about the zero value with a level of scatter of the order of the noise level in the data profile. Systematic features such as curvature usually signal the need for additional components, or modelling with inadequate functions, or biased data (caused by e.g., unmasked dust). While the addition of more components will invariably improve the fit, it must nevertheless remain physically motivated, i.e., the user should seek evidence for the presence of extra components, either in the image, ellipticity, PA or $B_4$ profile. As noted in Sec. 3.4 of \cite{Dullo&Graham2014}, one should not blindly add S\'ersic components.

The user can choose the radial range of data to be considered throughout the fit, by entering start and stop values (in arcsec). While excluding any data is not generally desirable (unless there are biasing factors in a particular range), particularly in the central regions, where most of the signal is, varying the radial extent of the data can provide an idea of the stability of the chosen model, and the uncertainties in its parameters.

\section{EXAMPLE DECOMPOSITIONS}\label{sec:ex}

\subsection{NGC~3348 -- A Cored Elliptical Galaxy}

In the first example I consider the galaxy NGC~3348, a massive elliptical galaxy with a cored inner SBP (\citealt{Rest+2001}, \citealt{GrahamEA2003}). Archival $HST$ data taken with the $ACS/WFC$ camera ($F814W$ filter) was retrieved from the {\it Hubble Legacy Archive}\footnote{\url{http://hla.stsci.edu}}. The sky background was measured with {\sc Imexamine} close to the chip edges, and subtracted from the image. The galaxy light profile was extracted from the resulting image with {\sc Isofit} and the PSF was characterised from the image by fitting a Moffat profile to bright stars, with {\sc Imexamine}.

The galaxy was modelled with a single core-S\'ersic component, in the range 0 -- 50 arcsec (roughly the distance from the photocentre to 3 out of 4 chip edges of the ACS/WFC chip), and the result is displayed in Figure \ref{fig:csersic}. The single-component fit yielded a core radius of $R_b = 0.43$ arcsec, break sharpness $\alpha = 1.86$, core slope $\gamma = 0.09$, half-light radius $R_e=27.63$ arcsec, and S\'ersic index $n=4.91$. These results are generally in good agreement with \cite{GrahamEA2003}, though the outer S\'ersic parameters, $R_e$ and $n$, are both $\sim 21\%$ higher in this analysis. The break radius agrees well with their reported value of $R_b=0.45$ arcsec, whereas the inner profile slope is shallower in this work than their reported value of $\gamma = 0.18$.

When interpreting these differences, it must be taken into consideration that this analysis was performed on imaging with different spatial resolution and at a longer wavelength (\citealt{GrahamEA2003} used $WFPC2/F555W$ data). Additionally, and perhaps more importantly, \cite{GrahamEA2003} performed the decomposition on deconvolved data from \cite{Rest+2001}, whereas \prof~accounts for seeing effects by convolving the model in stead. As \cite{Ferrarese+2006} point out, deconvolving (noisy) data can lead to unstable results, whereas the convolution of a noisless model is mathematically a more well defined process, and hence is more robust. \prof's convolution scheme was tested for S\'ersic and core-S\'ersic models and Gaussian seeing, by modelling synthetic images (with known light profiles) that were generated and convolved with independent software (the IRAF tasks {\sc Bmodel} and {\sc Gauss}).

\begin{figure}
\includegraphics[width=1.\columnwidth]{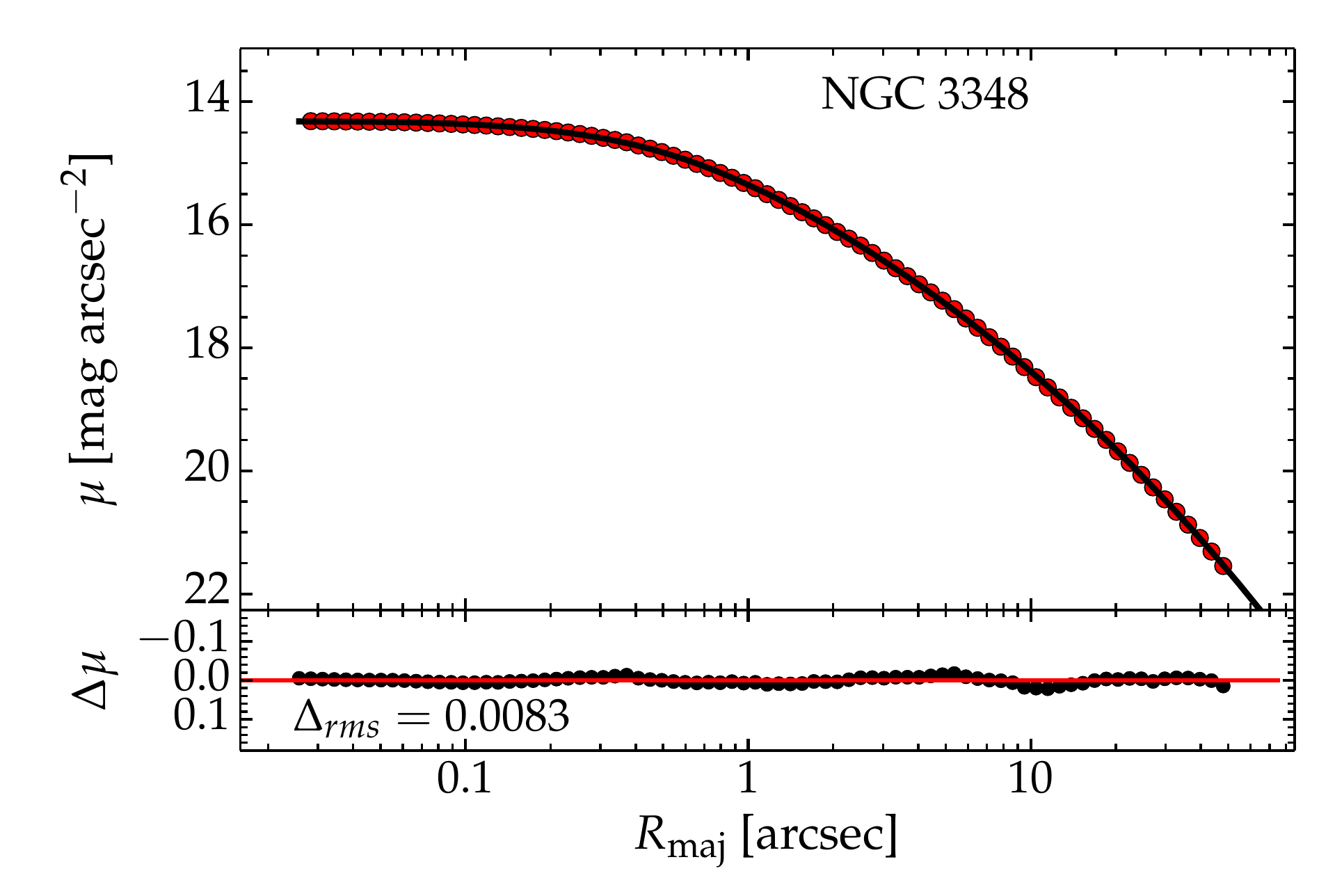}
\caption{Major axis light profile (red circles) of the cored elliptical galaxy NGC~3842 fit with a core-S\'ersic profile (black curve; Equation \ref{equ:core-sersic}), with break radius $R_b=0.43$ arcsec, inner slope $\gamma=0.09$ and break sharpness $\alpha=1.86$. The profile beyond $R_b$ has a S\'ersic index of $n=4.91$ and half-light radius of $R_e=27.63$ arcsec. Data from the $HST$, $F814W$ filter.}
\label{fig:csersic}
\end{figure}

\begin{figure*}
\includegraphics[width=1.\textwidth]{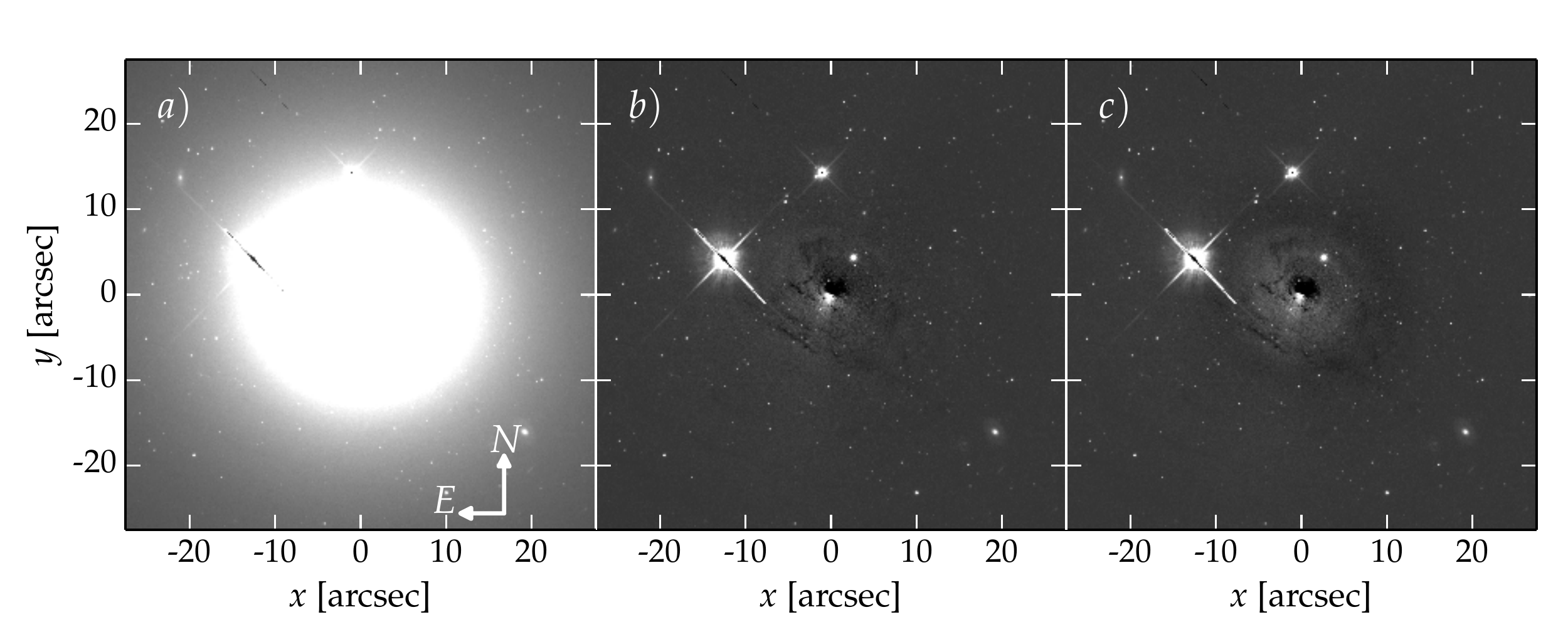}
\caption{{\it Panel $a)$:} HST ($F814W$) image of the cored early-type galaxy NGC~3348. {\it Panel $b)$:} Image in $a)$ minus a 2D reconstruction generated with {\sc Cmodel} (see \citealt{Ciambur2015}), based on isophote fitting with {\sc Isofit}. {\it Panel $c)$:} Image in $a)$ minus a  reconstruction based on the same isophote tables but with the data surface brightness column (red circles in Figure \ref{fig:csersic}, top panel) replaced by decomposition model obtained with \prof~(black curve in Figure \ref{fig:csersic}, top panel). The image stretch was adjusted to reveal low small-level systematics ($< 0.05$ mag arcsec$^{-2}$) causing the appearence of waves in a pond (and correspond to the curvature in the residual profile $\Delta\mu(R)$, also shown in Figure \ref{fig:csersic}, second panel from the top). The central systematic indicates that the core region is offset from the photometric centre of the external isophotes.}
\label{fig:n7619phot}
\end{figure*}

The core-S\'ersic model was tested with \prof~ on four additional cored galaxies (namely NGC~1016, NGC~3842, NGC~5982,  and NGC~7619) and compared with results from \cite{Dullo&Graham2012} and \cite{Dullo&Graham2014} (who, like \citealt{GrahamEA2003}, have used deconvolved profiles, from \citealt{Lauer+2005}). The core slopes obtained with \prof~ appeared to be systematically shallower than the literature values computed from the deconvolved data\footnote{Note, however, that these past studies performed the decomposition with $\alpha$ held fixed, whereas \prof~ allows this parameter to remain free. This aspect may influence the core profile slope $\gamma$.}. If this is indeed a systematic discrepancy and not simply a chance occurence in the five galaxies considered here, this issue would imply that literature measurements of the cores' light deficit are biased-low. A more comprehensive study on a larger sample of cored galaxies would be required to confirm this, which is however beyond the scope of this paper. 

\subsection{Pox~52 -- Using  the Data Vector PSF Option}\label{sec:p52}

The second example is intended to illustrate how, when diffraction effects are significant, even the Moffat approximation of the true PSF is inadequate and can lead to wrong results. This can be avoided with \prof~through the use of the data vector PSF feature.

The data chosen for demonstrating this feature was an $HST$ image of the nucleated dwarf Seyfert 1 galaxy Pox~52 (\citealt{Kunth+1987}, \citealt{Barth+2004},  \citealt{Thornton+2008}), observed with the $ACS/HRC$ camera in the $I-$band ($F814W$ filter).

\begin{figure*}
\includegraphics[width=1.\textwidth]{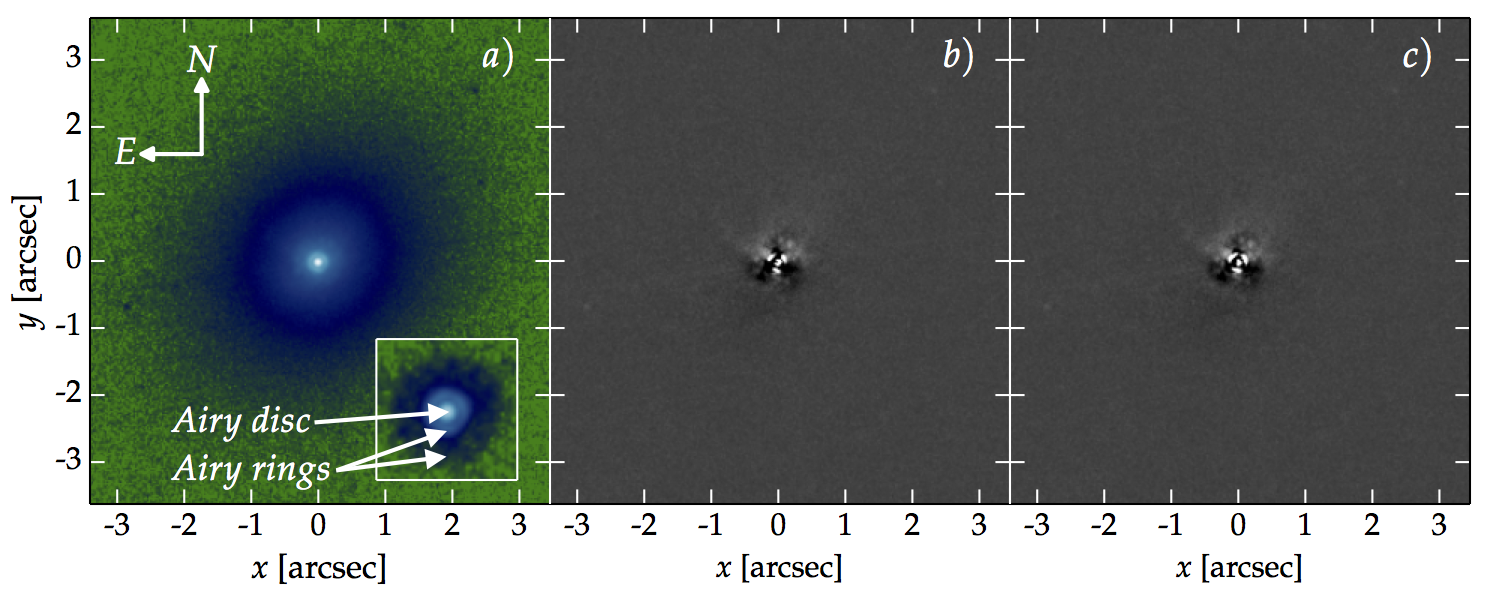}
\caption{$I-$band image of Pox~52, taken with the $ACS/HRC$ camera ($F814W$ filter) onboard the $HST$. The three panels are analogous to Figure \ref{fig:core-sersic}, except panel $a)$ is plotted on a logarithmic scale and false-colour scheme, for clarity. With a pixel size of 0.025 arcsec, the PSF is well sampled: the central point source displays a clear first Airy ring and a faint second. The Airy rings are also obvious in the surface brightness profile (Figure \ref{fig:num_psf}). The inset is a nearby bright star in the same data, to the SW of Pox~52. For clarity, it is zoomed-in by a ratio of 2:1 compared to the Pox~52 image.}
\label{fig:Pox52}
\end{figure*}

The bright point-source (AGN) at the centre is `spread' onto the detector into a distinct Airy pattern (see Figure \ref{fig:Pox52}), which is also obvious in the surface brightness profile (Figure \ref{fig:num_psf}). For this galaxy, the PSF was characterised from a bright, nearby star (inset of Figure \ref{fig:Pox52}), in two ways: $a)$ by fitting a Moffat profile with {\sc Imexamine} and $b)$ by fitting the star's light profile (extracted with {\sc Ellipse}) with four Gaussians (for the Airy disc and first three rings)\footnote{Note that a raw profile obtained with {\sc Ellipse} (or {\sc Isofit}) can be used as well, but this can be noisy at large distances from the star's centroid, so in this work this was modelled with 4 Gaussians, for a smooth result.}. The galaxy's SBP was then fit with \prof~with two components, namely a nuclear point source and a S\'ersic component. This was done for both PSF choices, and the results are displayed in Figure \ref{fig:num_psf}.

The best-fitting Moffat profile from {\sc Imexamine} had a FWHM of 3.04 pixels and a $\beta$ parameter of 7.41. The high value of $\beta$ indicates that {\sc Imexamine} fit essentially a Gaussian on just the Airy disc (first peak of the PSF) and ignored the wings (Airy rings).\footnote{This is probably caused by {\sc Imexamine}'s weighting scheme for pixels outside the half-maximum radius, which reduces the contribution of wings to the profile.}

\begin{figure*}
\includegraphics[width=1.\columnwidth]{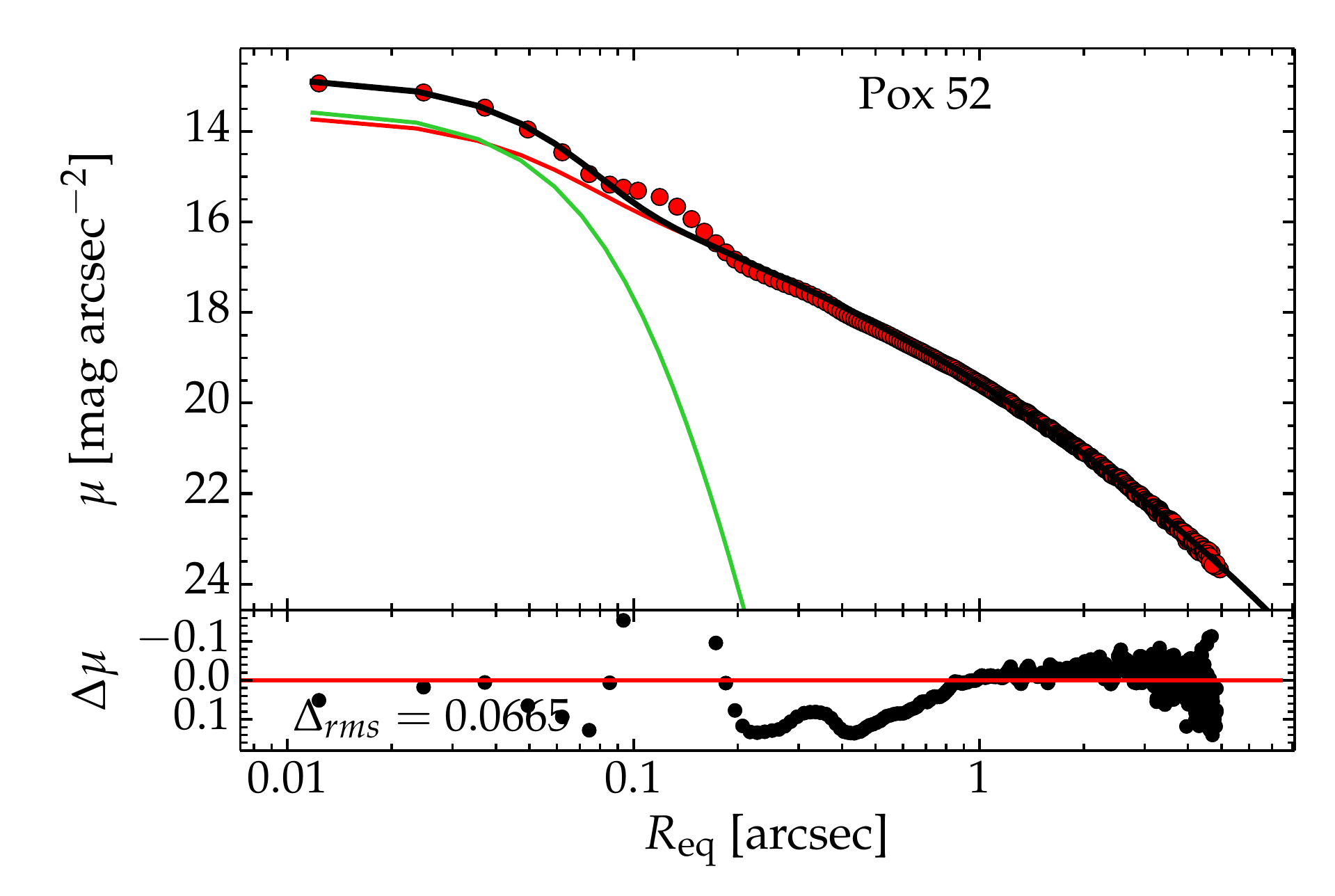}
\includegraphics[width=1.\columnwidth]{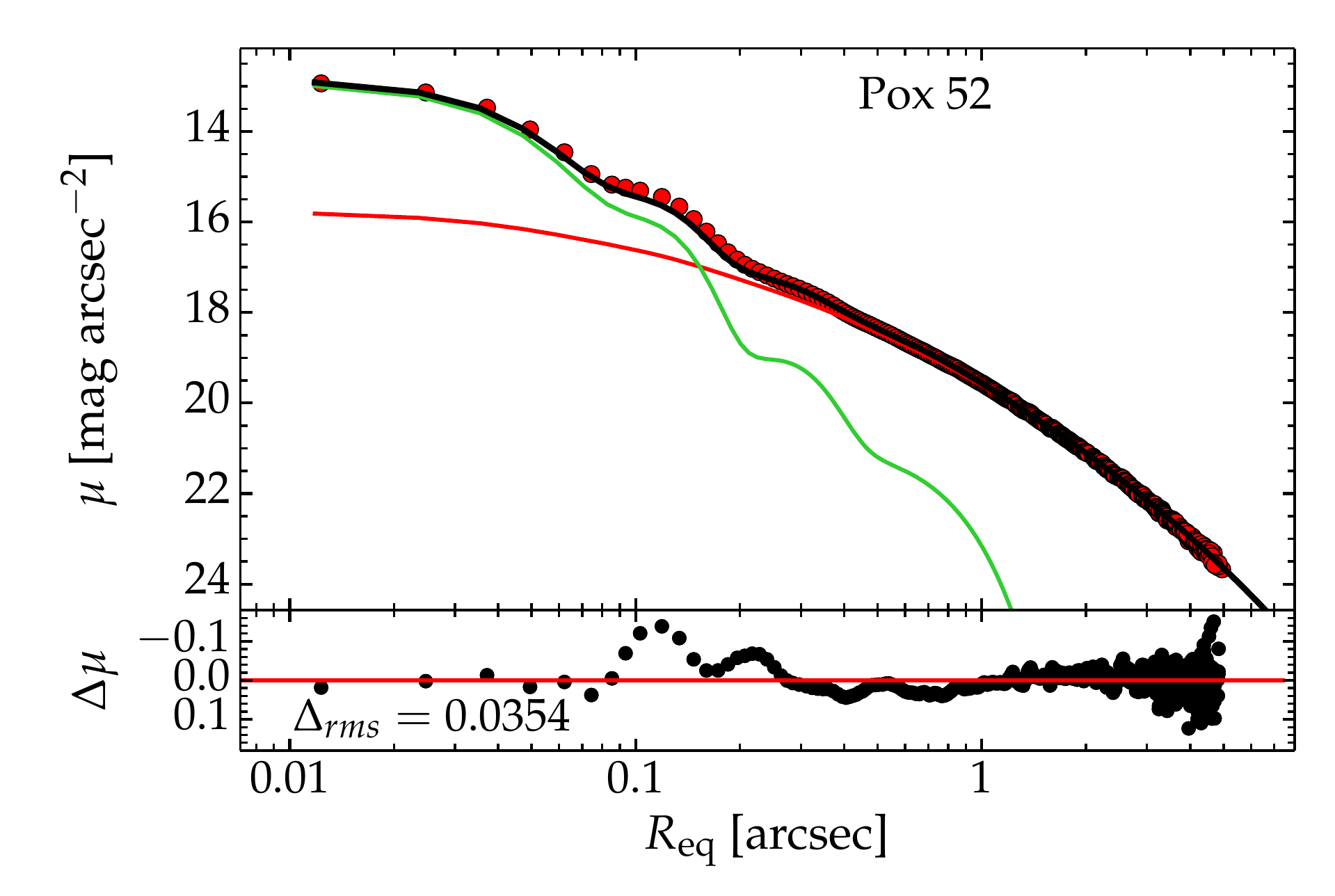}
\caption{Equivalent axis light profile of Pox~52 (red circles) modelled with two choices of PSF: a Moffat PSF (left-hand panel) and a data vector PSF (right-hand panel). The models (black curves) are each built from a point-source (green) and a S\'ersic component (red). The data vector PSF better captures the Airy rings (see Figure \ref{fig:Pox52}) and thus provides a superior model.}
\label{fig:num_psf}
\end{figure*}

The decomposition solution with the Moffat PSF is a 2-component model: a point source of central surface brightness $\mu_0 = 13.51$ and a S\'ersic component characterised by $\mu_e = 19.62$, $R_e = 1.03$ arcsec and $n = 4.19$. This solution is displayed in the left-hand panel of Figure \ref{fig:Pox52}. During the decomposition process, \prof~ tried to compensate for the unaccounted-for flux in the PSF wings (between 0.1 -- 0.3 arcsec) by making the S\'ersic component more concentrated than it should be. This illustrates a case when things went wrong, not because of {\sc Profiler} but because of the input PSF.

When performing the decomposition with a data vector PSF, the flux in the wings of the PSF is accounted for much more accurately, and the overall solution is better. Quantitatively, it was also a 2-component model, with the point source $\mu_0 = 12.92$ and the S\'ersic component $\mu_e = 20.11$, $R_e = 1.27$ arcsec and $n = 3.12$. The S\'ersic component is now less concentrated and its total magnitude $m = 16.33$ mag (in the Vega magnitude system) is $\sim$50\% fainter than in the previous case, but in good agreement with the value of 16.2 reported by \cite{Thornton+2008}. Additionally, the residual profile displays considerably reduced curvature beyond 0.1arcsec ($\Delta_{rms}$ is reduced by a factor of 2), though there is still systematic curvature at the scale of the first two Airy rings, which is due to the still imperfect PSF estimation.

\subsection{NGC~2549 -- One Spheroid, Two Bars and a Truncated Disc}\label{sec:p52}

\begin{figure*}
\includegraphics[width=1.\textwidth]{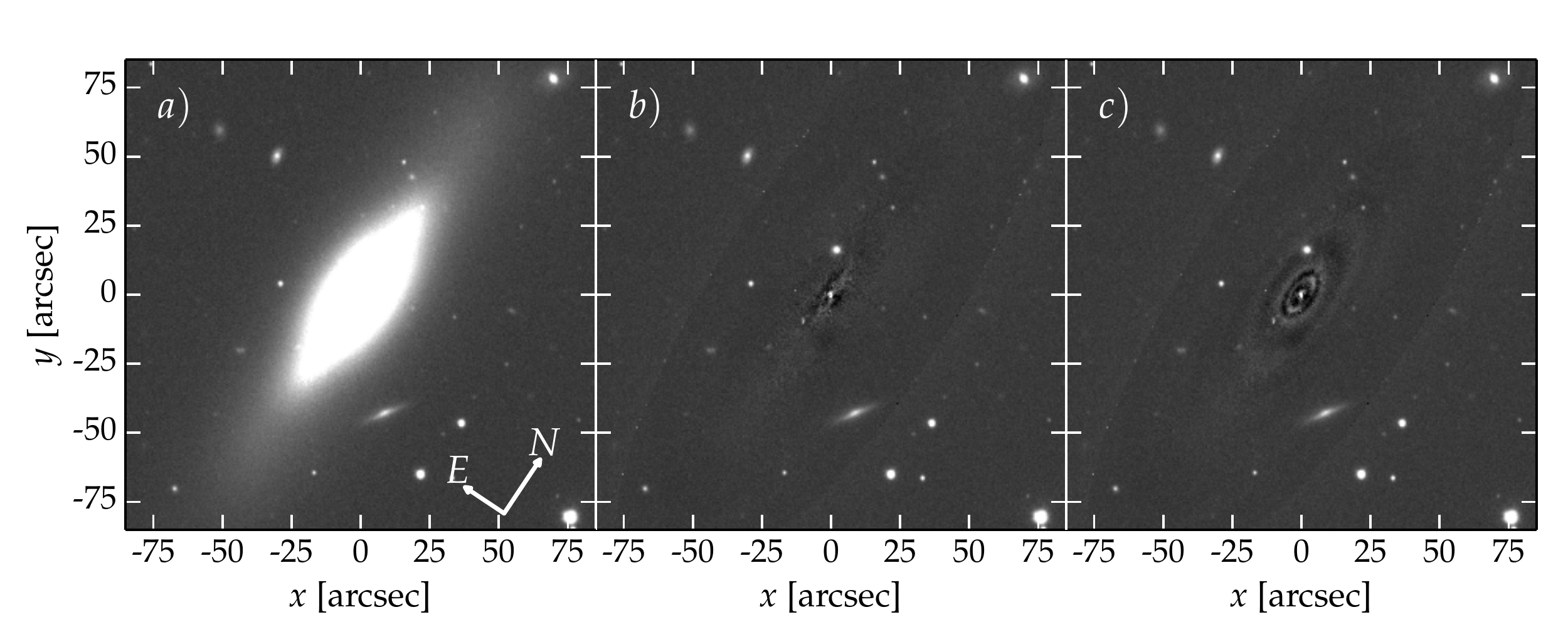}
\caption{{\it Panel $a)$:} SDSS $r-$band image of NGC~2549. {\it Panel $b)$:} Image in $a)$ minus a 2D reconstruction generated with {\sc Cmodel} (see \citealt{Ciambur2015}), based on isophote fitting with {\sc Isofit}. {\it Panel $c)$:} Image in $a)$ minus a 2D reconstruction with {\sc Cmodel}, based on the same isophote tables but with the data surface brightness column (red circles in Figure \ref{fig:n2549}, top panel) replaced by decomposition model obtained with \prof~(black curve in Figure \ref{fig:n2549}, top panel). The image stretch was adjusted to reveal low small-level systematics, which cause the appearence of waves in a pond (and correspond to the curvature in the residual profile $\Delta\mu(R)$, also shown in Figure \ref{fig:n2549}, second panel from the top). However, the nested peanut structures are well captured (there are no X-shaped systematics).}
\label{fig:n2549phot}
\end{figure*}

The third example involves the complex edge-on galaxy NGC~2549. Apart from a spheroid and a disc component, this object shows the signatures of two nested bars, and was shown to host nested X/peanut--shaped structures associated with the two bars (\citealt{Ciambur+2016}). 

SDSS $r-$band data from DR9 was analysed as before, and the best-fitting model consisted of a S\'ersic shperoid, two nested bars, also modelled with S\'ersic functions, and a truncated (Type II) exponential disc, with a break radius of 86.2 arcsec, inner scale length $h1=42.7$ arcsec and outer $h2=27.2$. The solution is displayed in Figure \ref{fig:n2549}, which also shows the ellipticity and $B_4$ harmonic profiles. Displaying these ancillary profiles is an option available to the user (as check-boxes in the GUI, see Figure \ref{fig:profiler}) and, in conjunction with the residual profile, they are often useful to signal the presence of additional components -- in this case, both $\epsilon(R_{\rm maj})$ and $B_4(R_{\rm maj})$ strongly indicate the presence of the inner bar component, and also display faint `bumps' corresponding to the outer bar, which is however more obvious in the SBP. The detection of the nested bars is particularly important given that this galaxy is viewed edge-on, i.e., the most difficult orientation for finding bars.

\begin{figure}
\includegraphics[width=1.\columnwidth]{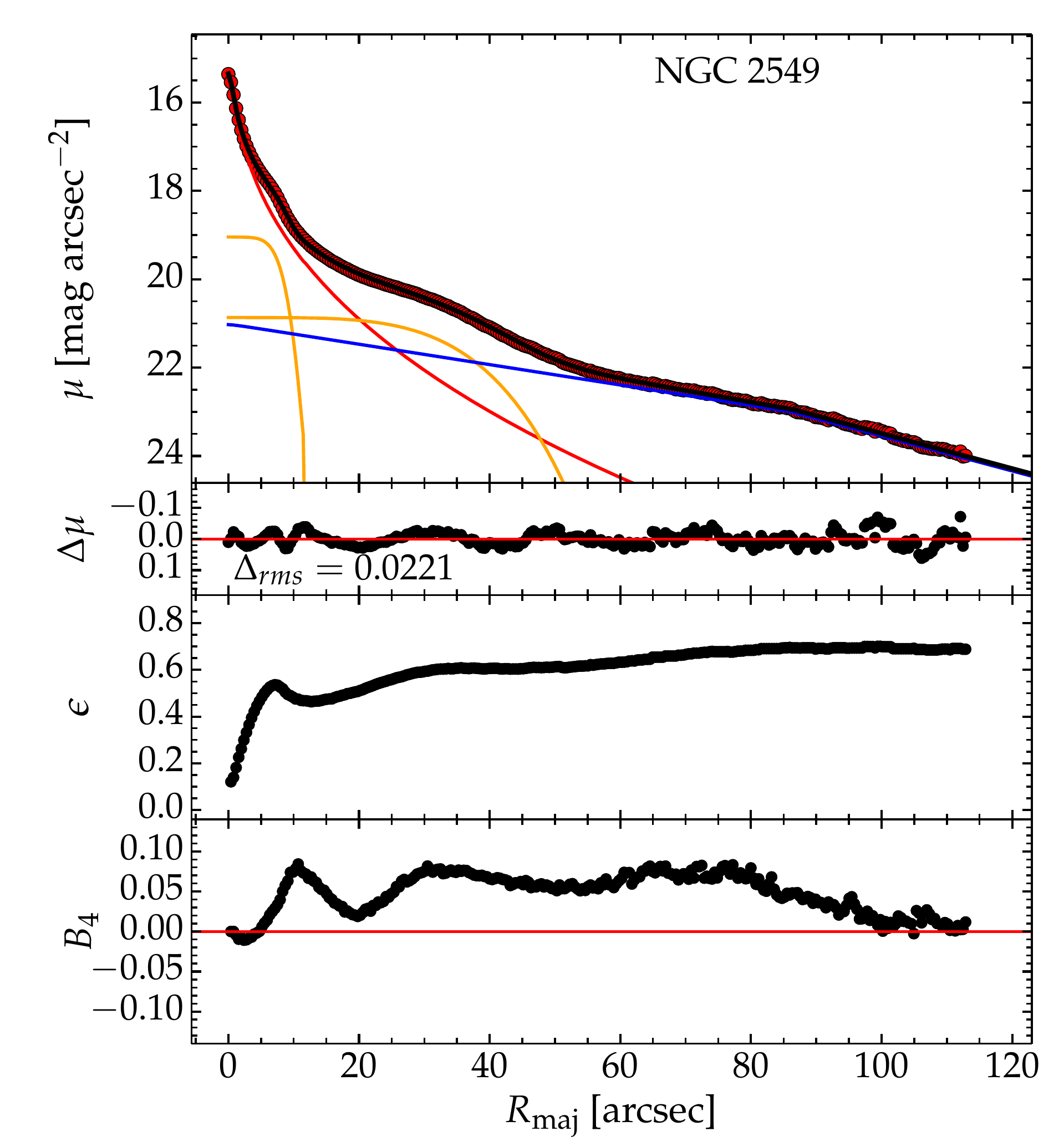}
\caption{Top panel shows the major axis decomposition of the edge-on, double-bar (nested peanut) galaxy NGC~2549, based on SDSS $r-$band data. The model is made up of a S\'ersic spheroid (red), two nested, S\'ersic bars (orange; $n=0.15$ for the inner, $n=0.23$ for the outer) and a truncated disc (blue) with a truncation radius of $R_b=86$ arcsec. Directly underneath is he residual profile, followed by the ellipticity ($\epsilon$) and $B_4$ (boxyness/discyness) profiles.}
\label{fig:n2549}
\end{figure}

As before, a 2D reconstruction of the galaxy image was performed with {\sc Cmodel}, based on the best-fitting SBP and the isophote parameters computed by {\sc Isofit}. This was substacted from the original galaxy image, resulting in a residual image displayed in Figure \ref{fig:n2549phot}, panel $(c)$. The appearence of waves in-a-pond reflects the curvature in the residual profile $\Delta\mu$ (Figure \ref{fig:n2549}). Note, however, that there are no X-shaped systematics, which indicates that the peanut bulges were well captured by the isophotal analysis.

\section{CONCLUSIONS}\label{sec:cons}

I have introduced \prof, a flexible and user-friendly program coded in {\sc Python}, designed to model radial surface brightness profiles of galaxies. 

With an intuitive GUI, \prof~can model a wide range of galaxy components, such as spheroids (elliptical galaxies or the bulges of spiral or lenticular galaxies), face-on, inclined, edge-on and (anti)truncated discs, resolved or un-resolved nuclear point-sources, bars, rings, spiral arms, etc. with an arsenal of analytical functions routinely used in the field, such as the S\'ersic, core-S\'ersic, exponential, edge-on disc model, Gaussian, Moffat and Ferrers' functions. In addition, \prof~can employ a broken exponential model (relevant for disc truncations or antitruncations) and two 1D special cases of the 2D edge-on disc model, namely along the major axis (in the disc plane) and along the minor axis (perpendicular to the disc plane).

\prof~is optimised to analyse isophote tables generated by the IRAF tasks {\sc Ellipse} and {\sc Isofit} but can also analyse two-column tables of radius and surface brightness. After reaching the best-fitting solution, the corresponding model parameters are returned. The major and equivalent axis profiles can both be analysed, and for the latter profile, each component's total magnitude is additionally returned.

The model convolution with the PSF is performed in 2D, with an FFT-based scheme. This allows for elliptical models, and additionally ensures that the convolution conserves the model's total flux (as a 1D convolution of the model profile with the PSF profile does not). Further, \prof~allows for a choice between Gaussian, Moffat or a user-provided data vector for the PSF (a table of $R$ and $I(R)$ values). All of the possible PSF choices can also be used as point-source components in the model.

\prof~is freely available from the following URL: \url{https://github.com/BogdanCiambur/PROFILER}.

\section{ACKNOWLEDGEMENTS}

I am grateful to  A. Graham for teaching me the subtleties of galaxy decomposition, and for reading parts of this manuscript. I also thank G. Savorgnan and B. Dullo for useful discussions.
Funding for SDSS-III has been provided by the Alfred P. Sloan Foundation, the Participating Institutions, the National Science Foundation, and the U.S. Department of Energy Office of Science. Part of this work is based on observations made with the NASA/ESA {\it Hubble Space Telescope}, and obtained from the {\it Hubble Legacy Archive}, which is a collaboration between the {\it Space Telescope Science Institute} (STScI/NASA), the {\it Space Telescope European Coordinating Facility} (ST-ECF/ESA) and the {\it Canadian Astronomy Data Centre} (CADC/NRC/CSA).

\bibliographystyle{mn2e}
\bibliography{references}

\begin{appendix}
\section{Appendix}

\begin{figure*}
\centering
\rotatebox{90}{\includegraphics[width=1.2\textwidth]{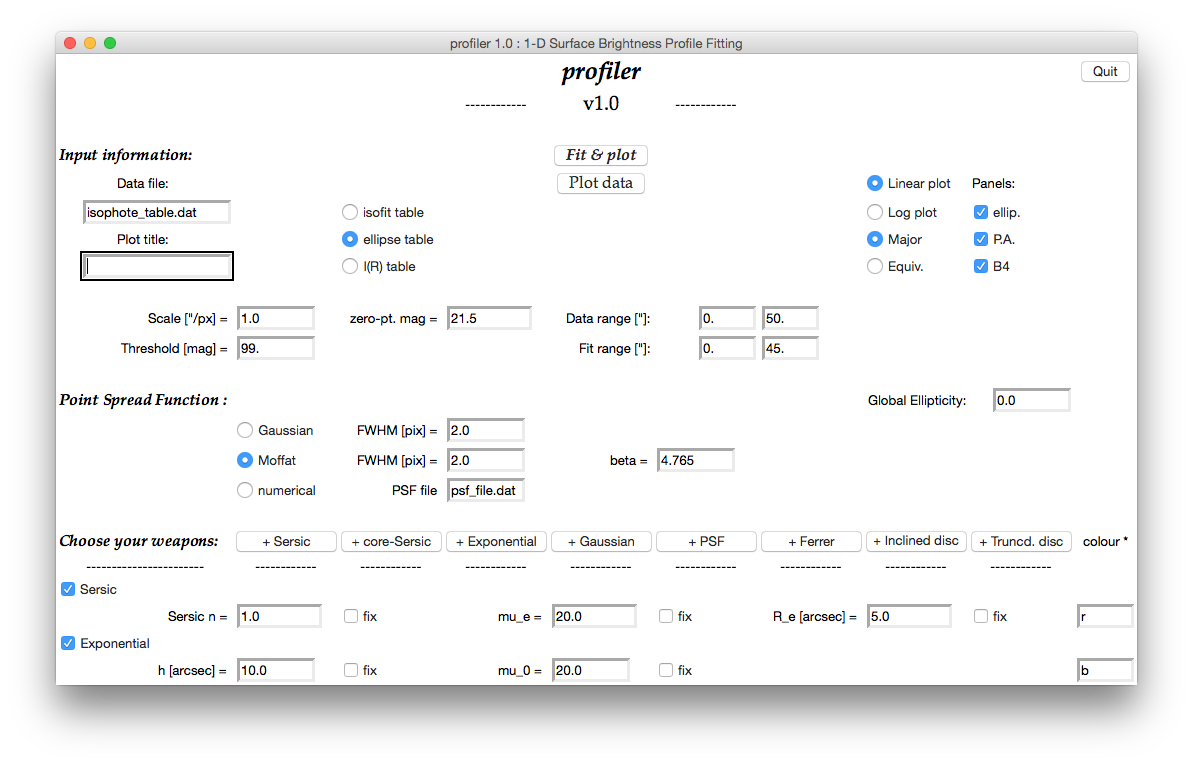}}
\caption{The {\sc Profiler} GUI, two active components (S\'ersic and exponential, added by clicking the corresponding buttons from the list above the two entries, for illustration purposes. All the text-boxes and check-boxes are set to default values, which the user must change to the specifics of the data (see text for details). The component parameters too must be set to initial guess-values, from which the code obtains the best-fitting solution.}
\label{fig:profiler}
\end{figure*}

\end{appendix}

\end{document}